\pdfoutput=1
\documentclass[11pt,a4paper]{article}
\usepackage{jcappub}
\usepackage[utf8]{inputenc}
\allowdisplaybreaks

\usepackage{lmodern}
\usepackage[T1]{fontenc}
\DeclareFontShape{OMX}{cmex}{m}{n}{
  <-7.5> cmex7
  <7.5-8.5> cmex8
  <8.5-9.5> cmex9
  <9.5-> cmex10
}{}
\SetSymbolFont{largesymbols}{normal}{OMX}{cmex}{m}{n}
\SetSymbolFont{largesymbols}{bold}  {OMX}{cmex}{m}{n}

\usepackage{amsmath}
\usepackage{amssymb}
\usepackage{mathtools}
\mathtoolsset{mathic} 

\DeclareMathAlphabet{\mathsfi}{OT1}{cmss}{m}{sl}
\DeclareMathAlphabet{\mathbfi}{OML}{cmm}{b}{it}
\DeclareSymbolFont{rsfs}{U}{rsfs}{m}{n}
\DeclareSymbolFontAlphabet{\mathscr}{rsfs}

\let\originalleft\left
\let\originalright\right
\renewcommand{\left}{\mathopen{}\mathclose\bgroup\originalleft}
\renewcommand{\right}{\aftergroup\egroup\originalright}

\newenvironment{equations}[1][]{\subequations\ifx\relax#1\relax\else\label{#1}\fi\align\ignorespaces}{\endalign\ignorespacesafterend\endsubequations}
\makeatletter
\def\@spliteq#1{\begin{equation}\begin{split}#1\end{split}\end{equation}}
\def\splitequation{\collect@body\@spliteq}

\makeatother

\usepackage{booktabs}

\usepackage{float}
\usepackage[labelformat=simple]{subcaption}

\usepackage{amsthm}

\newcommand{\eqend}[1]{\,\mathrm{#1}}

\newcommand{\laplace}{\mathop{}\!\bigtriangleup}

\newcommand{\supp}{\operatorname{supp}}
\newcommand{\hankel}[1]{\mathop{}\!\mathrm{H}^{(#1)}}

\renewcommand{\vec}[1]{{\ifnum9<1#1\mathbf{#1}\else\ifcat\noexpand#1\relax\boldsymbol{#1}\else\mathbfi{#1}\fi\fi}}

\newcommand{\mathe}{\mathrm{e}}
\newcommand{\mathi}{\mathrm{i}}
\newcommand{\total}{\mathop{}\!\mathrm{d}}

\makeatletter
\DeclareRobustCommand{\abs}{\bgroup\@ifstar\@@abs\@abs}
\newcommand*{\@@abs}[2][]{{#1\lvert{#2}#1\rvert}\egroup}
\newcommand*{\@abs}[1]{{\left\lvert{#1}\right\rvert}\egroup}
\DeclareRobustCommand{\norm}{\bgroup\@ifstar\@@norm\@norm}
\newcommand*{\@@norm}[2][]{{#1\lVert{#2}#1\rVert}\egroup}
\newcommand*{\@norm}[1]{{\left\lVert{#1}\right\rVert}\egroup}

\DeclareRobustCommand{\bra}{\bgroup\@ifstar\@@bra\@bra}
\newcommand*{\@@bra}[2][]{{#1\langle{#2}#1\vert}\egroup}
\newcommand*{\@bra}[1]{{\left\langle{#1}\right\vert}\egroup}
\DeclareRobustCommand{\ket}{\bgroup\@ifstar\@@ket\@ket}
\newcommand*{\@@ket}[2][]{{#1\vert{#2}#1\rangle}\egroup}
\newcommand*{\@ket}[1]{{\left\vert{#1}\right\rangle}\egroup}

\DeclareRobustCommand{\bigo}{\bgroup\@ifstar\@@bigo\@bigo}
\newcommand*{\@@bigo}[2][]{{\mathcal{O}#1({#2}#1)}\egroup}
\newcommand*{\@bigo}[1]{{\mathcal{O}\left({#1}\right)}\egroup}
\makeatother

\usepackage{xspace}

\newcommand{\ie}{\textit{i.\,e.}\xspace}
\newcommand{\eg}{\textit{e.\,g.}\xspace}

\newtheorem*{lemma*}{Lemma}
\renewcommand{\proof}{\hspace{-\parindent}\textbf{Proof.}\hspace{1em}}
\renewcommand{\endproof}{\hfill $\square$\smallskip\hspace{-\parindent}}

\begin{document}

\title{Compactly supported linearised observables in single-field inflation}

\author[1]{Markus B. Fr\"ob,}
\emailAdd{mbf503@york.ac.uk}

\author[2]{Thomas-Paul Hack,}
\emailAdd{thomas-paul.hack@itp.uni-leipzig.de}

\author[1]{Atsushi Higuchi}
\emailAdd{atsushi.higuchi@york.ac.uk}

\affiliation[1]{Department of Mathematics, University of York, Heslington, York, YO10 5DD, United Kingdom}
\affiliation[2]{Institut f{\"u}r Theoretische Physik, Universit{\"a}t Leipzig, Br{\"u}derstra{\ss}e 16, 04103 Leipzig, Germany}

\abstract{We investigate the gauge-invariant observables constructed by smearing the graviton and inflaton fields by compactly supported tensors at linear order in general single-field inflation. These observables correspond to gauge-invariant quantities that can be measured locally. In particular, we show that these observables are equivalent to (smeared) local gauge-invariant observables such as the linearised Weyl tensor, which have better infrared properties than the graviton and inflaton fields. Special cases include the equivalence between the compactly supported gauge-invariant graviton observable and the smeared linearised Weyl tensor in Minkowski and de~Sitter spaces. Our results indicate that the infrared divergences in the tensor and scalar perturbations in single-field inflation have the same status as in de~Sitter space and are both a gauge artefact, in a certain technical sense, at tree level.}

\keywords{quantum field theory on curved space, cosmological perturbation theory}

\maketitle

\section{Introduction}

Recently inflationary cosmological models~\cite{kazanas_apj_1980,sato_mnras_1981,guth_prd_1981,linde_plb_1982,albrecht_steinhardt_prl_1982} have been studied extensively since experimental evidence has been accumulating to support inflation (see, \eg, the recent observations reported in~\cite{planck2013_i,planck2013_xvi,planck2015_xx}). In these models, the graviton and inflaton fields, or tensor and scalar perturbations, suffer from infrared (IR) divergences~\cite{ford_parker_prd_1977a,ford_parker_prd_1977b}. However, the physical significance of those divergences  has been debated for some time, and various methods to deal with them have been proposed~\cite{tsamis_woodard_npb_1996,tsamis_woodard_ap_1997,garriga_tanaka_prd_2008,tsamis_woodard_prd_2008,janssen_prokopec_cqg_2008,riotto_sloth_jcap_2008,burgess_et_al_jcap_2010,rajaraman_kumar_leblond_prd_2010,seery_cqg_2010,urakawa_tanaka_prd_2010,kahya_onemli_woodard_plb_2011,giddings_sloth_jcap_2011,gerstenlauer_et_al_jcap_2011,giddings_sloth_prd_2011,garbrecht_rigopoulos_prd_2011,xue_gao_brandenberger_jcap_2012,pimentel_senatore_zaldarriaga_jhep_2012,senatore_zaldarriaga_jhep_2013,assassi_baumann_green_jhep_2013,urakawa_tanaka_cqg_2013,tanaka_urakawa_ptep_2014}. Although most researchers argue that these IR divergences are not observable at tree level and that the problems only arise at higher order, it is recognised that the IR divergences of the tree-level correlators of the tensor and scalar perturbations are responsible for the IR problems at higher order. Therefore, it will be useful to thoroughly understand the nature of the IR divergences, and the related issues of long-distance and late-time (secular) growth of the two-point correlators for the tensor and scalar perturbations at linear order.

An important issue in understanding IR divergences in a generally covariant theory, such as inflationary cosmological models, is gauge invariance. Only gauge-invariant quantities can be observed, and for this reason they are often called simply \emph{observables}. Since physical observations are mostly local, it is useful to discuss observables that are defined in a compact spacetime region. Although the metric and other perturbations themselves are not gauge invariant, one can construct compactly supported and gauge-invariant observables by smearing them by compactly supported functions, or test functions, satisfying certain conditions. An observable of this type has been used recently to discuss quantisation of pure linearised gravity in spacetimes with a positive cosmological constant~\cite{fewster_hunt_rmp_2013}, as well as for the quantisation of the linearised graviton-inflaton system on arbitrary backgrounds, including those with FLRW symmetry~\cite{hack_cqg_2014}.

In this paper we study similar observables in single-field inflation, obtained by smearing the graviton and inflaton fields by test functions which satisfy a condition ensuring gauge invariance. In particular, we show that these observables are equivalent to a collection of smeared local gauge-invariant fields, defined at each spacetime point\footnote{From now on we use the word ``local'' to mean ``defined at each spacetime point'' unless otherwise stated.}, such as the linearised Weyl tensor. In other words, we show that the graviton and inflaton fields smeared in a gauge-invariant manner contain exactly the same information as certain local gauge-invariant fields. These local fields have better IR behaviour than the graviton and inflaton fields upon quantisation. We find that, as a result, the compactly supported gauge-invariant graviton-inflaton quantum correlators are IR finite, decay for large spacelike separations, and have no secular growth at late times. For the special cases of Minkowski and de~Sitter spaces, the graviton smeared in a gauge-invariant manner is equivalent to the smeared linearised Weyl tensor, as already shown in~\cite{higuchi_arxiv_2012}.

The rest of the paper is organised as follows: in section~\ref{review}, we review linear perturbations in single-field inflation and present some formulae useful later, referring to appendix~\ref{appendix_conformal}, which presents some facts related to the conformal transformation. In section~\ref{equivalence}, we show that the observables obtained by smearing the graviton and inflaton fields in a gauge-invariant manner are equivalent to certain (smeared) local gauge-invariant fields by using two theorems established in appendices~\ref{appendix_lemma} and~\ref{appendix_onshell_scalar}. In section~\ref{correlators}, we express these local gauge-invariant fields in terms of the tensor perturbation and the Sasaki--Mukhanov variable, which describes the scalar perturbation. In section~\ref{ir_behaviour} we examine the IR finiteness of the gauge-invariant graviton-inflaton quantum correlation functions, and in section~\ref{section:power_spectra} we 
give their relation to the scalar and tensor power spectra.  Finally, we conclude the paper in section~\ref{discussion} with a discussion of some of our results. In appendix~\ref{appendix_lemma2} we give a proof of a well-known Poincar{\'e} lemma for compactly supported divergence-free vector fields, and appendix~\ref{appendix_geometric} presents an attempt to express the local gauge-invariant fields mentioned above as a linear perturbation of geometric quantities. Our metric signature is mostly plus and we use natural units with $\hbar = c = 1$.

\section{Linearised perturbations in single-field inflation}
\label{review}

The single-field inflationary model is described by the metric $\tilde{g}_{\mu\nu}$ and the inflaton field $\tilde{\phi}$, and its action is given by
\begin{equation}
S = \frac{1}{\kappa^2} \int \tilde{R} \sqrt{-\tilde{g}} \total^n x - \frac{1}{2} \int
\left[ \tilde{\nabla}^\mu \tilde{\phi} \tilde{\nabla}_\mu \tilde{\phi} + V(\tilde{\phi}) \right] \sqrt{-\tilde{g}} \total^n x \eqend{.}
\end{equation}
A cosmological constant can be incorporated into the potential $V(\tilde\phi)$. We obtain the Einstein equations as the Euler--Lagrange equation for $\tilde{g}_{\mu\nu}$:
\begin{equation}
\label{emunu_def}
\tilde{E}_{\mu\nu} \equiv 2 \tilde{R}_{\mu\nu} - \tilde{g}_{\mu\nu} \tilde{R} - \kappa^2 \tilde{T}_{\mu\nu} = 0 \eqend{,}
\end{equation}
where the stress tensor is given by
\begin{equation}
\label{tmunu_def}
\tilde{T}_{\mu\nu} \equiv \tilde{\nabla}_\mu \tilde{\phi} \tilde{\nabla}_\nu \tilde{\phi} - \frac{1}{2} \tilde{g}_{\mu\nu} \left( \tilde{\nabla}^\rho \tilde{\phi} \tilde{\nabla}_\rho \tilde{\phi} + V(\tilde{\phi}) \right) \eqend{.}
\end{equation}
The Euler-Lagrange equation for $\tilde{\phi}$ is the scalar field equation,
\begin{equation}
\label{f_def}
\tilde{F} \equiv \tilde{\nabla}^\mu \tilde{\nabla}_\mu \tilde{\phi} - \frac{1}{2} V'(\tilde{\phi}) = 0 \eqend{.}
\end{equation}
The background metric $g_{\mu\nu}$ is conformally flat and, thus, takes the form,
\begin{equation}
g_{\mu\nu} = a^2 \eta_{\mu\nu}\eqend{,}
\end{equation}
where $\eta_{\mu\nu}$ is the flat metric and the scale factor $a(\eta)$ depends only on conformal time $\eta$.

Let us discuss some properties of the background fields that will be useful later. The Hubble and deceleration parameters are defined by
\begin{equations}[H_and_epsilon_def]
H &\equiv \frac{a'}{a^2} \eqend{,} \\
\epsilon &\equiv - \frac{H'}{H^2 a} \eqend{,}
\end{equations}
respectively, where a prime denotes the derivative with respect to the conformal time $\eta$. We also use later the other slow-roll parameter defined by
\begin{equation}
\label{delta_def}
\delta \equiv \frac{\epsilon'}{2 H a \epsilon} \eqend{.}
\end{equation}
The background Riemann tensor can be found using equation~\eqref{app_riemann} as
\begin{equation}
R_{\mu\rho\nu\sigma} = H^2 \left[ \left( g_{\mu\nu} g_{\rho\sigma} - g_{\mu\sigma} g_{\nu\rho} \right) + \epsilon \left( t_\mu t_\nu g_{\rho\sigma} - t_\rho t_\nu g_{\mu\sigma} - t_\mu t_\sigma g_{\rho\nu} + t_\rho t_\sigma g_{\mu\nu} \right) \right] \eqend{,}
\end{equation}
where $t^{\mu} \equiv a^{-1} (\partial/\partial \eta)^\mu$ is the future-directed unit vector on the background spacetime. The Ricci and scalar curvatures can readily be found from this formula as
\begin{equations}
R_{00} &= - (n-1) (1-\epsilon) H^2 a^2 \eqend{,} \\
R_{kl} &= (n-1-\epsilon) H^2 a^2 \delta_{kl} \eqend{,} \\
R &= (n-1) (n-2\epsilon) H^2 \eqend{,}
\end{equations}
where $k,l,\ldots$ are space indices. Then, from the space and time components of the Einstein equations~\eqref{emunu_def} for the background fields one finds
\begin{equations}[eom_background_phi2]
\kappa^2 V(\phi) &= 2 (n-2) (n-1-\epsilon) H^2 \eqend{,} \\
\kappa^2 (\phi')^2 &= 2 (n-2) \epsilon H^2 a^2 \eqend{.}
\end{equations}
The background scalar field $\phi$ satisfies the following field equation resulting from~\eqref{f_def}:
\begin{equation}
\label{eom_background_phi}
\phi'' + (n-2) H a \phi' + \frac{1}{2} a^2 V'(\phi) = 0 \eqend{.}
\end{equation}

We consider perturbations around this background, with the full metric $\tilde{g}_{\mu\nu}$ and scalar field $\tilde{\phi}$ decomposed as
\begin{equations}[linear_exp]
\tilde{g}_{\mu\nu} &= g_{\mu\nu} + g^{(1)}_{\mu\nu} \eqend{,} \\
\tilde{\phi} &= \phi + \phi^{(1)} \eqend{,}
\end{equations}
and linearise the field equations~\eqref{emunu_def} and \eqref{f_def}. The linear parts of the equations $\tilde{F} = 0$ and $\tilde{E}_{\mu\nu} = 0$ are
\begin{equation}
\label{eom_linear_f}
F^{(1)} \equiv \left( \nabla^\mu \nabla_\mu - \frac{1}{2} V''(\phi) \right) \phi^{(1)} - \left( \nabla_\mu g_{(1)}^{\mu\nu} - \frac{1}{2} \nabla^\nu g^{(1)} \right) \nabla_\nu \phi - g_{(1)}^{\mu\nu} \nabla_\mu \nabla_\nu \phi = 0
\end{equation}
and
\begin{splitequation}
\label{eom_linear_e}
E^{(1)}_{\mu\nu} &\equiv 2 \nabla^\rho \nabla_{(\mu} g^{(1)}_{\nu)\rho} - \nabla^\rho \nabla_\rho g^{(1)}_{\mu\nu} - \nabla_\mu \nabla_\nu g^{(1)} - g_{\mu\nu} \left( \nabla^\rho \nabla^\sigma g^{(1)}_{\rho\sigma} - \nabla^\rho \nabla_\rho g^{(1)} \right) \\
&\quad+ g_{(1)}^{\rho\sigma} \left( R_{\rho\sigma} g_{\mu\nu} - g_{\rho\mu} g_{\sigma\nu} R \right) - \kappa^2 T^{(1)}_{\mu\nu} = 0 \eqend{,}
\end{splitequation}
where the linear part of the stress tensor is given by
\begin{splitequation}
T^{(1)}_{\mu\nu} &= 2 \nabla_{(\mu} \phi \nabla_{\nu)} \phi^{(1)} - \frac{1}{2} g^{(1)}_{\mu\nu} \left( \nabla^\rho \phi \nabla_\rho \phi + V(\phi) \right) \\
&\quad- \frac{1}{2} g_{\mu\nu} \left( 2 g^{\rho\sigma} \nabla_\rho \phi \nabla_\sigma \phi^{(1)} - g_{(1)}^{\rho\sigma} \nabla_\rho \phi \nabla_\sigma \phi + V'(\phi) \phi^{(1)} \right) \eqend{.}
\end{splitequation}
In these equations the indices are raised and lowered by the background metric $g_{\mu\nu}$, and the covariant derivative $\nabla_\mu$ is compatible with it: $\nabla_\mu g_{\nu\lambda} = 0$.

The linearised gauge transformation of any tensor $\tilde{X}_{\alpha\beta\cdots\gamma}$ with respect to the vector $\xi^\mu$ is its Lie derivative with respect to $\xi^\mu$, which we denote by $\mathscr{L}_\xi\tilde{X}_{\alpha\beta\cdots\gamma}$. Thus, to first order we have
\begin{equations}[gauge_trafo]
\delta g^{(1)}_{\mu\nu} &= \mathscr{L}_\xi g_{\mu\nu} = 2 \nabla_{(\mu} \xi_{\nu)} \eqend{,} \\
\delta \phi^{(1)} &= \mathscr{L}_\xi \phi = \xi^\alpha \nabla_\alpha \phi \eqend{.}
\end{equations}
Suppose now that $\tilde{X}_{\alpha\beta\cdots\gamma}$ is a tensor constructed from $\tilde{g}_{\mu\nu}$ and $\tilde{\phi}$. Suppose further that, upon substitution of~\eqref{linear_exp}, it can be written as
\begin{equation}
\tilde{T}_{\alpha\beta\cdots\gamma} = T_{\alpha\beta\cdots\gamma} + T^{(1)}_{\alpha\beta\cdots\gamma}
\end{equation}
to first order in $g_{\mu\nu}^{(1)}$ and $\phi^{(1)}$, where $T_{\alpha\beta\cdots\gamma}$ and $T^{(1)}_{\alpha\beta\cdots\gamma}$ are of zeroth and first order in these fields, respectively. Then, under the gauge transformation~\eqref{gauge_trafo} one has~\cite{stewart_walker_prsl_1974}
\begin{equation}
\label{gauge_trafo_lie}
\delta T_{\alpha\beta\cdots\gamma}^{(1)} = \mathscr{L}_\xi T_{\alpha\beta\cdots\gamma} \eqend{.}
\end{equation}
Thus $T_{\alpha\beta\cdots\gamma}^{(1)}$ is gauge invariant if $\mathscr{L}_\xi T_{\alpha\beta\cdots\gamma} = 0$, which in particular is the case if the background value vanishes: $T_{\alpha\beta\cdots\gamma} = 0$. In particular, $\delta F^{(1)} = \mathscr{L}_\xi F$ and $\delta E^{(1)}_{\mu\nu} = \mathscr{L}_\xi E_{\mu\nu}$, and hence
the linear quantities $F^{(1)}$ and $E^{(1)}_{\mu\nu}$ are gauge invariant if the background equations, $E_{\mu\nu} = 0$ and $F = 0$, are satisfied, which we will assume from now on. Later on, a Bianchi identity for these quantities will also be useful, which we obtain by linearising the identity
\begin{equation}
\tilde{\nabla}^\mu \tilde{E}_{\mu\nu} = - \kappa^2 \tilde{\nabla}^\mu \tilde{T}_{\mu\nu} = - \kappa^2 \tilde{F} \tilde{\nabla}_\nu \tilde{\phi}
\end{equation}
[following from the usual Bianchi identity $\tilde{\nabla}^\mu \left( 2 \tilde{R}_{\mu\nu} - \tilde{g}_{\mu\nu} \tilde{R} \right) = 0$] and using the background field equations $E_{\mu\nu} = 0$ and $F = 0$. This results in
\begin{equation}
\label{eom_conserved}
\nabla^\mu E^{(1)}_{\mu\nu} = - \kappa^2 F^{(1)} \nabla_\nu \phi \eqend{,}
\end{equation}
which is identically satisfied for \emph{arbitrary} perturbations $g_{\mu\nu}^{(1)}$ and $\phi^{(1)}$.

\section{Equivalence of the compactly supported graviton-inflaton observables and local gauge-invariant observables}
\label{equivalence}

The graviton and inflaton perturbations, $g_{\mu\nu}^{(1)}(x)$ and $\phi^{(1)}(x)$, become singular upon quantisation. For example, their two-point correlation functions (correlators) are divergent when the two points coincide. For this reason, it is convenient to consider the quantities obtained by smearing them over a compact region in spacetime. Thus we define
\begin{equation}
\label{i_def}
I(f,\chi) \equiv \int \left( f^{\mu\nu} g^{(1)}_{\mu\nu} + 2 \chi \phi^{(1)} \right) \sqrt{-g} \total^n x \eqend{,}
\end{equation}
where $f^{\mu\nu}$ and $\chi$ are smooth and of compact support, and $f^{\mu\nu}$ is symmetric. (Tensors will be assumed to be smooth from now on unless otherwise stated.) In fact, quantities of this type are the fundamental objects in the algebraic formulation of quantum field theory~\cite{haag_kastler_jmp_1964,haag1992}. These objects have an additional advantage that they can readily be made gauge invariant (under linear gauge transformations) without sacrificing locality, unlike the original fields $g_{\mu\nu}^{(1)}(x)$ and $\phi^{(1)}(x)$. To see this, we note that the change of $I(f,\chi)$ under the gauge transformation~\eqref{gauge_trafo} is given by
\begin{splitequation}
\delta I(f,\chi) &= 2 \int \left( f^{\mu\nu} \nabla_\mu \xi_\nu + \chi \xi^\mu \nabla_\mu \phi \right) \sqrt{-g} \total^n x \\
&= 2 \int \xi_\nu \left( - \nabla_\mu f^{\mu\nu} + \chi \nabla^\nu \phi \right) \sqrt{-g} \total^n x \eqend{,}
\end{splitequation}
where the surface term in the second line is absent because $f^{\mu\nu}$ is compactly supported. Thus, the integral $I(f,\chi)$ is a gauge-invariant observable whenever the following condition is satisfied:
\begin{equation}
\label{fmunu_chi_gaugeinv_cond}
\nabla_\mu f^{\mu\nu} = \chi \nabla^\nu \phi \eqend{.}
\end{equation}
We impose this condition from now on so that $I(f,\chi)$ is gauge invariant. This observable (with $\chi = 0$) was used in~\cite{fewster_hunt_rmp_2013} to discuss quantisation of linearised gravity in general Einstein spacetimes with a positive cosmological constant. The analysis of ref.~\cite{fewster_hunt_rmp_2013} on the basis of $I(f,\chi)$ was later generalised to the linearised graviton-inflaton system on arbitrary on-shell backgrounds, including those with FLRW symmetry~\cite{hack_cqg_2014}.

Although we motivated the observable $I(f,\chi)$ as a regularised version of local fields, after imposing the condition~\eqref{fmunu_chi_gaugeinv_cond} it is impossible to take the limit where the compactly supported functions $f^{\mu\nu}$ and $\chi$ become local, \ie, a constant tensor multiplying a Dirac $\delta$ distribution supported at a point. If such a limit were possible, the observable $I(f,\chi)$ would reduce to a linear combination of $g^{(1)}_{\alpha\beta}$ and $\phi^{(1)}$, which cannot be gauge invariant, contradicting the gauge invariance of $I(f,\chi)$. One might argue that a gauge-invariant graviton field $g^{(1)}_{\alpha\beta}$ can be defined by fixing the gauge completely, but in general a complete gauge fixing involves field transformations which are not compactly supported. For example, to extract the usual gauge-invariant Bardeen potentials $\Phi_\text{A/H}$ from a general metric perturbation one has to solve the equations\footnote{These equations can be derived from the decomposition~\eqref{h_decomp} of the metric perturbation for general $X^\mu$, noting that $S = 2 \Phi_\text{A}$ and $\Sigma = 2 \Phi_\text{H}$.}
\begin{equations}
\laplace^2 \Phi_\text{A} &= \frac{1}{2} \laplace^2 h_{00} - \left( \partial_\eta + H a \right) \laplace \partial^k h_{0k} + \frac{1}{2 (n-2)} \left( \partial_\eta + H a \right) \partial_\eta \left[ (n-1) \partial^k \partial^l h_{kl} - \delta^{kl} \laplace h_{kl} \right] \eqend{,} \\
\begin{split}
\laplace^2 \Phi_\text{H} &= \frac{1}{2 (n-2)} \delta^{kl} \laplace^2 h_{kl} - \frac{1}{2 (n-2)} \laplace \partial^k \partial^l h_{kl} + H a \laplace \partial^k h_{0k} \\
&\quad- \frac{(n-1)}{2 (n-2)} H a \partial^k \partial^l h_{kl}' + \frac{1}{2 (n-2)} H a \delta^{kl} \laplace h_{kl}' \eqend{,}
\end{split}
\end{equations}
where $\partial^k \equiv \delta^{kl} \partial_l$, $\laplace \equiv \partial^k \partial_k$, and the field $h_{\mu\nu}$ is the rescaled perturbation defined by
\begin{equation}
\label{hmunu_def}
h_{\mu\nu} \equiv a^{-2} g^{(1)}_{\mu\nu} \eqend{,}
\end{equation}
which is usually employed in cosmology. Therefore, unless the right-hand sides are of the form $\laplace^2 f$ for some compactly supported function $f$ (which will not be the case in general even if $h_{\mu\nu}$ has compact support), the Bardeen potentials will be non-local. Nevertheless, we will argue at the end of this section that a certain local limit of $I(f,\chi)$ does exist, although it is different from the na\"{\i}ve one.

Writing the gauge invariance condition~\eqref{fmunu_chi_gaugeinv_cond} as
\begin{equation}
\label{flat_rep}
\frac{1}{a^{n+2}} \partial_\mu \left( a^{n+2} f^{\mu\nu} \right) + \frac{1}{a} t^\nu \left( H a f + \phi' \chi \right) = 0 \eqend{,}
\end{equation}
where $f \equiv g_{\mu\nu} f^{\mu\nu}$, we can readily find some solutions to this equation. Let $u^{\alpha\mu\beta\nu}$ be any compactly supported tensor which is antisymmetric in the first and last two indices, and symmetric under the pairwise exchange $(\alpha\mu) \leftrightarrow (\beta\nu)$\footnote{Since $f^{\mu\nu}$ in equation~\eqref{particular_solution} does not depend on the part of $u^{\alpha\mu\beta\nu}$ totally antisymmetric in
the last three indices, we may also require $u^{\alpha[\mu\beta\nu]} = 0$. In fact, the $u^{\alpha\mu\beta\nu}$ we construct in appendix~\ref{appendix_lemma} has this property.}. Then a solution to equation~\eqref{flat_rep} can be given as
\begin{equations}[particular_solution]
f^{\mu\nu} & = a^{-n-2} \partial_\alpha \partial_\beta u^{\alpha\mu\beta\nu} \eqend{,} \\
\chi &= - \frac{H a}{\phi'} f \eqend{.}
\end{equations}
The main technical result of this paper is that this is actually the general solution \emph{on-shell}, \ie, \emph{any} observable $I(f,\chi)$ can be represented with $f^{\mu\nu}$ and $\chi$ of the form~\eqref{particular_solution} if the fields $g^{(1)}_{\mu\nu}$ and $\phi^{(1)}$ satisfy the field equations $F^{(1)} = 0$~\eqref{eom_linear_f} and $E_{\mu\nu}^{(1)} = 0$~\eqref{eom_linear_e}. Moreover, the support of the tensor $u^{\alpha\mu\beta\nu}$ can be chosen to be close to the support of the original $f^{\mu\nu}$ and $\chi$ in the technical sense explained in appendix~\ref{appendix_lemma}. We first explain this result and its consequences for pure linearised gravity in Minkowski and de~Sitter spaces~\cite{higuchi_arxiv_2012}.

\subsection{The Minkowski case}

We let $\chi = 0$ and consider
\begin{equation}
I_0(f) \equiv \int f^{\mu\nu} g^{(1)}_{\mu\nu} \sqrt{-g} \total^n x \eqend{.}
\end{equation}
(Here we have $\sqrt{-g} = 1$ in Cartesian coordinates, of course.) The condition~\eqref{fmunu_chi_gaugeinv_cond} for the gauge invariance of $I_0(f)$ becomes $\partial_\mu f^{\mu\nu} = 0$. Thus, we want to show that a compactly supported divergence-free symmetric tensor $f^{\mu\nu}$ can always be expressed as
$f^{\mu\nu} = \partial_\alpha \partial_\beta u^{\alpha\mu\beta\nu}$, where the compactly supported tensor $u^{\alpha\mu\beta\nu}$ has the symmetry properties and support stated before. This result is proved in appendix~\ref{appendix_lemma}; it is related to the so-called Calabi complex~\cite{khavkine_jgeomphys_2014}, and a slightly different proof can be found in ref.~\cite{higuchi_arxiv_2012}. It is analogous to the compact-support version of the Poincar{\'e} lemma for $p$-forms: on an $n$-dimensional Euclidean space a closed $p$-form of compact support with $1 \leq p \leq n-1$ is exact. Thus, in Minkowski space we have
\begin{equation}
I_0(f) = - \frac{1}{2} \int u^{\alpha\mu\beta\nu} R^{(1)}_{\alpha\mu\beta\nu} \total^n x \eqend{,}
\end{equation}
where
\begin{equation}
R^{(1)}_{\alpha\mu\beta\nu} \equiv \partial_\alpha \partial_{[\nu} g^{(1)}_{\beta]\mu} - \partial_\mu \partial_{[\nu} g^{(1)}_{\beta]\alpha}
\end{equation}
is the linearised Riemann tensor. It is traceless because the linearised Ricci tensor vanishes on-shell in linearised gravity in Minkowski space. Thus, we have
on-shell
\begin{equation}
I_0(f) = - \frac{1}{2} \int u^{\alpha\mu\beta\nu} C^{(1)}_{\alpha\mu\beta\nu} \total^n x \eqend{,}
\end{equation}
where $C^{(1)}_{\alpha\mu\beta\nu}$ is the linearised Weyl tensor, which is the traceless part of $R^{(1)}_{\alpha\mu\beta\nu}$. That is, the graviton observable $I_0(f)$ is equivalent to the smeared linearised Weyl tensor.

\subsection{The de~Sitter case}

The de~Sitter case is achieved by choosing $V(\phi) = 2 \kappa^{-2} (n-2) (n-1) H^2$, with $H$ a positive constant, and $\phi' = 0$ [see equation~\eqref{eom_background_phi2}]. As in the Minkowski case, we consider the pure graviton observable $I_0(f)$, with $\nabla_\mu f^{\mu\nu} = 0$. A useful observation, which will be generalised in the next subsection, is that the observable $I_0(f)$ is invariant on-shell [\ie, when $E^{(1)}_{\mu\nu}\left( g^{(1)} \right) = 0$] under the transformation $f^{\mu\nu} \to \check{f}^{\mu\nu} \equiv f^{\mu\nu} + \delta f^{\mu\nu}$ with $\delta f^{\mu\nu} = E^{(1)\mu\nu}(v)$, where $v^{\mu\nu}$ is any
compactly supported tensor. This holds because
\begin{splitequation}
I_0(\check{f}) &= I_0(f) + \int E^{(1)\mu\nu}(v) g^{(1)}_{\mu\nu} \sqrt{-g} \total^n x \\
&= I_0(f) + \int v^{\mu\nu} E^{(1)}_{\mu\nu}\left( g^{(1)} \right) \sqrt{-g} \total^n x \\
&= I_0(f) \eqend{.}
\end{splitequation}
Note also that $\nabla_\mu \check{f}^{\mu\nu} = \nabla_\mu E^{(1)\mu\nu}(v) = 0$, as follows from the Bianchi identity~\eqref{eom_conserved} taking into account that $\phi = 0$ in de~Sitter. Now, this equation can be written as [see equation~\eqref{flat_rep}]
\begin{equation}
\label{ds_checkf_gi}
\partial_\mu \left( a^{n+2} \check{f}^{\mu\nu} \right) + a^{n+2} t^\nu H \check{f} = 0 \eqend{.}
\end{equation}
Hence, if the tensor $\check{f}^{\mu\nu}$ is traceless ($\check{f} = 0$), then $a^{n+2} \check{f}^{\mu\nu}$ will be divergence free with respect to the flat metric, and the result of appendix~\ref{appendix_lemma} can be used. We now show that it is possible to choose $\delta f^{\mu\nu}$ so that $\check{f}^{\mu\nu} = f^{\mu\nu} + \delta f^{\mu\nu}$ is traceless.

We find from equation~\eqref{eom_linear_e} that
\begin{splitequation}
E^{(1)}_{\mu\nu}(v) &\equiv 2 \nabla_{(\mu} \nabla^\rho v_{\nu)\rho} - \nabla^\rho \nabla_\rho v_{\mu\nu} - \nabla_\mu \nabla_\nu v - g_{\mu\nu} \left( \nabla^\rho \nabla^\sigma v_{\rho\sigma} - \nabla^\rho \nabla_\rho v \right) \\
&\quad+ (n-3) H^2 g_{\mu\nu} v + 2 H^2 v_{\mu\nu} \eqend{,}
\end{splitequation}
where we changed the order of some derivatives using $R_{\alpha\beta\gamma\delta} = H^2 \left( g_{\alpha\gamma} g_{\beta\delta} - g_{\alpha\delta} g_{\beta\gamma} \right)$. The trace reads
\begin{equation}
g^{\mu\nu} E_{\mu\nu}^{(1)}(v) = (2-n) \left( \nabla^\rho \nabla^\sigma v_{\rho\sigma} - \nabla^\rho \nabla_\rho v \right) + (n-1) (n-2) H^2 v \eqend{.}
\end{equation}
Thus by choosing
\begin{equation}
v_{\mu\nu} = \frac{1}{(n-2) H^2} \left( f_{\mu\nu} - \frac{1}{n-1} g_{\mu\nu} f \right) \eqend{,}
\end{equation}
and using $\nabla_\mu f^{\mu\nu} = 0$, we find $\delta f = g^{\mu\nu} E^{(1)}_{\mu\nu}(v) = - f$. Thus, we have $\check{f} = f + \delta f = 0$, and $\nabla_\mu \check{f}^{\mu\nu} = 0$ as shown above. From equation~\eqref{ds_checkf_gi}, we find that $\partial_\mu \left( a^{n+2} \check{f}^{\mu\nu} \right) = 0$, and hence by the results of appendix~\ref{appendix_lemma} there exists a compactly supported tensor $u^{\alpha\mu\beta\nu}$ with the symmetries of the Riemann tensor such that
\begin{equation}
\label{af_by_u}
a^{n+2} \check{f}^{\mu\nu} = \partial_\alpha \partial_\beta u^{\alpha\mu\beta\nu} \eqend{.}
\end{equation}
Hence
\begin{equation}
\label{If}
I_0(f) = I_0(\check{f}) = - \frac{1}{2} \int u^{\alpha\mu\beta\nu} \left( \partial_\alpha \partial_{[\nu} h_{\beta]\mu} - \partial_\mu \partial_{[\nu} h_{\beta]\alpha} \right) \total^n x \eqend{,}
\end{equation}
with the rescaled perturbation $h_{\mu\nu}$ defined by equation~\eqref{hmunu_def}.

The tensor $\partial_\alpha \partial_{[\nu}h_{\beta]\mu} - \partial_\mu \partial_{[\nu}h_{\beta]\alpha}$ is gauge invariant, but it is not the linearised Weyl tensor even on-shell because its trace does not vanish. In order to express the observable $I_0(f)$ in terms of the linearised Weyl tensor, we first separate out the traceless part $w^{\alpha\mu\beta\nu}$ of $u^{\alpha\mu\beta\nu}$ as
\begin{equation}
\label{u_decomposition}
u^{\alpha\mu\beta\nu} = w^{\alpha\mu\beta\nu} + \frac{2}{n-2} \eta^{\alpha[\beta} u^{\nu]\mu} - \frac{2}{n-2} \eta^{\mu[\beta} u^{\nu]\alpha}
+ \frac{2}{(n-1) (n-2)} \eta^{\alpha[\beta} \eta^{\nu]\mu} u
\end{equation}
with
\begin{equation}
u^{\mu\nu} \equiv \eta_{\alpha\beta} u^{\alpha\mu\beta\nu} \eqend{,} \qquad u \equiv \eta_{\mu\nu} u^{\mu\nu} \eqend{.}
\end{equation}
Substitution of equation~\eqref{u_decomposition} into \eqref{af_by_u} yields
\begin{equation}
\label{f_delta_f}
a^{n+2}\check{f}^{\mu\nu}
= \partial_\alpha \partial_\beta w^{\alpha\mu\beta\nu} + \frac{1}{n-2}s^{\mu\nu}\eqend{,}
\end{equation}
where
\begin{equation}
s^{\mu\nu} = \partial_\rho\partial^\rho u^{\mu\nu} - 2\partial^{(\mu} \partial_\rho u^{\nu)\rho} + \frac{1}{n-1}\left( \partial^\mu\partial^\nu u -
\eta^{\mu\nu} \partial_\rho \partial^\rho u\right)\eqend{.}
\end{equation}
Here indices are raised and lowered by $\eta_{\mu\nu}$. Next we show that the second term of \eqref{f_delta_f} can be written as
$\partial_\alpha\partial_\beta q^{\alpha\mu\beta\nu}$, where $q^{\alpha\mu\beta\nu}$ has the same symmetries as $w^{\alpha\mu\beta\nu}$ and
is traceless.

Since $0 = a^{n+2} \check{f} = \eta_{\mu\nu} \partial_\alpha \partial_\beta u^{\alpha\mu\beta\nu} = \partial_\alpha\partial_\beta u^{\alpha\beta}$, the tensor
$u^{\mu\nu}$ can be expressed as
\begin{equation}
u^{\mu\nu} = \partial_\rho \left( W^{\mu\rho\nu} + W^{\nu\rho\mu} \right) \eqend{,}
\end{equation}
where $W^{\mu\rho\nu}$ is antisymmetric in the last two indices and has support close to the support of $u^{\mu\nu}$, as shown in appendix~\ref{appendix_lemma2}. Define
\begin{equation}
K^{\alpha\mu\beta\nu} \equiv \frac{1}{n-3} \left( \partial^\beta W^{\nu\alpha\mu} - \partial^\nu W^{\beta\alpha\mu} + \partial^\alpha W^{\mu\beta\nu} - \partial^\mu W^{\alpha\beta\nu} \right) \eqend{.}
\end{equation}
The tensor $K^{\alpha\mu\beta\nu}$ is antisymmetric in the first two and last two indices, and symmetric under the exchange of these pairs of indices. Define $q^{\alpha\mu\beta\nu}$ to be the traceless part of this tensor. Then a straightforward calculation shows that
\begin{equation}
\partial_\alpha \partial_\beta q^{\alpha\mu\beta\nu} = \frac{1}{n-2} s^{\mu\nu} \eqend{.}
\end{equation}
Hence, defining a traceless tensor by
\begin{equation}
\check{u}^{\alpha\mu\beta\nu} \equiv w^{\alpha\mu\beta\nu} + q^{\alpha\mu\beta\nu} \eqend{,}
\end{equation}
we have from equation~\eqref{f_delta_f} that
\begin{equation}
a^{n+2} \check{f}^{\mu\nu} = \partial_\alpha \partial_\beta \check{u}^{\alpha\mu\beta\nu} \eqend{.}
\end{equation}
By substituting this expression into equation~\eqref{If} we obtain
\begin{splitequation}
\label{If2}
I_0(f) &= - \frac{1}{2} \int \check{u}^{\alpha\mu\beta\nu} \left( \partial_\alpha \partial_{[\nu} h_{\beta]\mu} - \partial_\mu \partial_{[\nu} h_{\beta]\alpha} \right) \total^n x \\
&= - \frac{1}{2} \int \check{u}^{\alpha\mu\beta\nu} C^{(1)\text{flat}}_{\alpha\mu\beta\nu} \total^n x \eqend{,}
\end{splitequation}
where $C^{(1)\text{flat}}_{\alpha\mu\beta\nu}$ is the linearised Weyl tensor in flat space with metric perturbation $h_{\mu\nu}$. This follows from the fact that $C^{(1)\text{flat}}_{\alpha\mu\beta\nu}$ is the traceless part of the linearised Riemann tensor $\partial_\alpha \partial_{[\nu} h_{\beta]\mu} - \partial_\mu \partial_{[\nu} h_{\beta]\alpha}$, and that $\check{u}^{\alpha\mu\beta\nu}$ is traceless.

Now, the full Weyl tensor $\tilde{C}_{\alpha\mu\beta\nu}$ for the metric $\tilde{g}_{\mu\nu} = a^2 \left( \eta_{\mu\nu} + h_{\mu\nu} \right)$ is related to the Weyl tensor $\tilde{C}^{\text{flat}}_{\alpha\mu\beta\nu}$ for the metric $\eta_{\mu\nu} + h_{\mu\nu}$ as $\tilde{C}_{\alpha\mu\beta\nu} = a^2 \tilde{C}^{\text{flat}}_{\alpha\mu\beta\nu}$ (see, \eg,~\cite{wald_textbook}). Since they both vanish for the background metric, the linearised Weyl tensors are also related in this manner:
\begin{equation}
\label{C_dS_flat}
C^{(1)}_{\alpha\mu\beta\nu} = a^2 C^{(1)\text{flat}}_{\alpha\mu\beta\nu} \eqend{.}
\end{equation}
By substituting this relation into equation~\eqref{If2}, we finally obtain
\begin{equation}
I_0(f) = - \frac{1}{2} \int \hat{u}^{\alpha\mu\beta\nu} C^{(1)}_{\alpha\mu\beta\nu} \sqrt{-g} \total^n x \eqend{,}
\end{equation}
where $\hat{u}^{\alpha\mu\beta\nu} \equiv a^{-(n+2)} \check{u}^{\alpha\mu\beta\nu}$ since $\sqrt{-g} = a^n$. Thus, the compactly supported graviton observable $I_0(f)$ in de~Sitter space is equivalent to the linearised Weyl tensor, which is gauge invariant because the background Weyl tensor vanishes~\cite{stewart_walker_prsl_1974}.

\subsection{The general case}

Now we study the observable $I(f,\chi)$ in a general FLRW spacetime assuming $\phi' \neq 0$. The test functions $f^{\mu\nu}$ and $\chi$ in the observable $I(f,\chi)$~\eqref{i_def}, which are compactly supported by definition, satisfy the equation~\eqref{flat_rep}. [Recall that this condition guarantees the gauge invariance of $I(f,\chi)$.] We consider the transformation $f^{\mu\nu} \to \check{f}^{\mu\nu} \equiv f^{\mu\nu} + \delta f^{\mu\nu}$, $\chi \to \check{\chi} \equiv \chi + \delta \chi$ with
\begin{equations}
\delta f^{\mu\nu} &= E^{(1)\mu\nu}(v,w) \eqend{,}\\
\delta \chi &= - \kappa^2 F^{(1)}(v,w) \eqend{,}
\end{equations}
where $E^{(1)\mu\nu}(v,w)$ and $F^{(1)}(v,w)$ are obtained by substituting $g^{(1)}_{\mu\nu} \to v_{\mu\nu}$ and $\phi^{(1)} \to w$ into $E^{(1)\mu\nu}$ and $F^{(1)}$ in equations~\eqref{eom_linear_e} and~\eqref{eom_linear_f}, respectively. The observable $I(f,\chi)$ is invariant under this transformation on-shell since
\begin{splitequation}
I(\check{f},\check{\chi}) &= I(f,\chi) + \int \left[ E^{(1)}_{\mu\nu}\left( v, w \right) g_{(1)}^{\mu\nu} - 2 \kappa^2 F^{(1)}\left( v, w \right) \phi^{(1)} \right]
\sqrt{-g} \total^n x \\
&= I(f,\chi) + \int \left[ v^{\mu\nu} E^{(1)}_{\mu\nu}\left( g^{(1)}, \phi^{(1)} \right) - 2 \kappa^2 w F^{(1)}\left( g^{(1)}\eqend{,}
\phi^{(1)} \right) \right] \sqrt{-g} \total^n x \\
&= I(f,\chi) \eqend{.}
\end{splitequation}
From the background Bianchi identity~\eqref{eom_conserved}
\begin{equation}
\nabla^\mu E^{(1)}_{\mu\nu}(v,w) = - \kappa^2 F^{(1)}(v,w) \nabla_\nu \phi \eqend{,}
\end{equation}
which holds for any $v^{\mu\nu}$ and $w$, we obtain $\nabla^\mu \delta f_{\mu\nu} = \delta \chi \nabla_\nu \phi$. Together with the gauge invariance condition~\eqref{fmunu_chi_gaugeinv_cond}, this implies that $\nabla_\mu \check{f}^{\mu\nu} = \check{\chi} \nabla^\nu \phi$, which we write in the form
\begin{equation}
\label{check_f_eq}
\frac{1}{a^{n+2}} \partial_\mu \left( a^{n+2} \check{f}^{\mu\nu} \right) + \frac{1}{a} t^\nu \left( H a \check{f} + \phi' \check{\chi} \right) = 0 \eqend{.}
\end{equation}
In appendix~\ref{appendix_onshell_scalar} we show that if we choose
\begin{equations}[w_v_expressions]
w & = A(\eta) \frac{2 (n-1) (n-2) \chi + \kappa \sqrt{2 (n-2) \epsilon} f}{2 \kappa^2 \left[ (n-2) (n-1-\epsilon) + 2 (n-1) \delta \right]} \eqend{,} \label{w_expression} \\
v^{\mu\nu} & = A(\eta) \left( f^{\mu\nu} - \frac{1}{n-1} g^{\mu\nu} f \right) + \frac{\phi'}{H a} g^{\mu\nu} w
\end{equations}
with
\begin{equation}
\label{def_A_eta}
A(\eta) = \frac{a}{(n-2) H} (\eta_0 - \eta) \eqend{,}
\end{equation}
where $\eta_0$ is any constant, then $H a \check{f} + \phi' \check{\chi} = 0$. This and equation~\eqref{check_f_eq} imply
\begin{equations}[check_equations]
&\partial_\mu \left( a^{n+2} \check{f}^{\mu\nu} \right) = 0 \eqend{,} \\
&\check{\chi} = - \frac{H a}{\phi'} \check{f} \eqend{.} \label{check_eq_2}
\end{equations}
We emphasise that this is an exact result for an arbitrary potential $V(\phi)$, and that no approximation is necessary for it to be valid.
Then, by appendix~\ref{appendix_lemma} there is a tensor $u^{\alpha\mu\beta\nu}$ with the symmetry of the Riemann tensor such that
\begin{equation}
\check{f}^{\mu\nu} = a^{-(n+2)} \partial_\alpha \partial_\beta u^{\alpha\mu\beta\nu} \eqend{.}
\end{equation}

Since the tensor $\check{f}^{\mu\nu}$ is not necessarily traceless this time, the observable $I(f,\chi) = I\left( \check{f}, \check{\chi} \right)$ cannot be reduced to the linearised Weyl tensor alone. To identify the Weyl-tensor component we separate out the traceless part $w^{\alpha\mu\beta\nu}$ of $u^{\alpha\mu\beta\nu}$ as in equation~\eqref{u_decomposition}. Then it follows that on-shell, if the field equations~\eqref{eom_linear_e} and~\eqref{eom_linear_f} for $g^{(1)}_{\mu\nu}$ and $\phi^{(1)}$ are satisfied, we have, noting equation~\eqref{check_eq_2},
\begin{splitequation}
\label{i_equiv_weyl_1}
I(f,\chi) &= \int a^{n+2} f^{\mu\nu} \left( h_{\mu\nu} - \frac{2 H a}{\phi'} \eta_{\mu\nu} \phi^{(1)} \right) \total^n x  \\
& = \int u^{\alpha\mu\beta\nu} \partial_\alpha \partial_\beta \left( h_{\mu\nu} - \frac{2 H a}{\phi'} \eta_{\mu\nu} \phi^{(1)} \right) \total^n x \\
&= - \frac{1}{2} \int w^{\alpha\mu\beta\nu} C^{(1)\text{flat}}_{\alpha\mu\beta\nu} \total^n x - \int u^{\mu\nu} C_{\mu\nu} \total^n x \eqend{,}
\end{splitequation}
where the flat-space linearised Weyl tensor $C^{(1)\textrm{flat}}_{\alpha\mu\beta\nu}$ is defined in the previous subsection and where
\begin{splitequation}
\label{C_expression}
C_{\mu\nu} &\equiv \frac{1}{n-2} \left( 2 \partial_{(\mu} \partial^\beta h_{\nu)\beta} - \partial^2 h_{\mu\nu} - \partial_\mu \partial_\nu h \right) - \frac{1}{(n-1) (n-2)} \eta_{\mu\nu} \left( \partial^\alpha \partial^\beta h_{\alpha\beta} - \partial^2 h \right) \\
&\quad+ 2 \partial_\mu \partial_\nu \left( \frac{H a}{\phi'} \phi^{(1)} \right) \eqend{.}
\end{splitequation}
The last equality in equation~\eqref{i_equiv_weyl_1} follows because $w^{\alpha\mu\beta\nu}$ is traceless, and equation~\eqref{C_dS_flat} holds here as well. Hence by defining
\begin{equations}
\hat{w}^{\alpha\mu\beta\nu} &\equiv a^{-(n+2)} w^{\alpha\mu\beta\nu} \eqend{,} \\
\hat{u}^{\mu\nu} &\equiv a^{-n} u^{\mu\nu} \eqend{,}
\end{equations}
we find
\begin{equation}
\label{final_equivalence}
I(f,\chi) = - \frac{1}{2} \int \hat{w}^{\alpha\mu\beta\nu} C_{\alpha\mu\beta\nu}^{(1)} \sqrt{-g} \total^n x - \int \hat{u}^{\mu\nu} C_{\mu\nu} \sqrt{-g} \total^n x \eqend{.}
\end{equation}

The linearised Weyl tensor $C^{(1)}_{\alpha\mu\beta\nu}$ is gauge invariant as in the de~Sitter case because the background Weyl tensor vanishes. The gauge invariance of $C_{\mu\nu}$ follows from that of $I(f,\chi)$, but can also be checked explicitly. For this, note that the gauge transformation can be expressed as
\begin{equations}[gauge_trafo_perts]
\delta h_{\mu\nu} &= 2 \partial_{(\mu} \xi_{\nu)} - 2 H a \eta_{\mu\nu} \xi_0 \eqend{,} \\
\delta \phi^{(1)} & = - \phi' \xi_0 \eqend{.}
\end{equations}
Interestingly, the tensor $C_{\mu\nu}$ is invariant under two separate transformations $\delta_1$ and $\delta_2$, given by $\delta_1 h_{\mu\nu} = 2 \partial_{(\mu} \xi_{\nu)}$, $\delta_1 \phi^{(1)} = 0$ and $\delta_2 h_{\mu\nu} = - 2 H a \eta_{\mu\nu} \xi_0$, $\delta_2 \phi^{(1)} = - \phi' \xi_0$, respectively.  We have attempted to express the tensor $C_{\mu\nu}$ as the linear perturbation of a geometric quantity, and have succeeded in doing so except for the component $C_{00}$, as shown in appendix~\ref{appendix_geometric}.

We have argued before that it does not make sense to consider local (Dirac $\delta$) limits of the test tensors $f^{\mu\nu}$ and $\chi$ in $I(f,\chi)$, because of the differential gauge-invariance constraint~\eqref{fmunu_chi_gaugeinv_cond} which they have to satisfy. However, given $I(f,\chi)$ in the equivalent form~\eqref{final_equivalence}, one may now consider local limits of the test tensors $\hat{w}^{\alpha\mu\beta\nu}$ and $\hat{u}^{\mu\nu}$, \ie, one can consider the limiting case where these test tensors are given by constant tensors multiplying a Dirac $\delta$ distribution supported at a point. From these $\hat{w}^{\alpha\mu\beta\nu}$ and $\hat{u}^{\mu\nu}$ one can obtain the corresponding $f^{\mu\nu}$ and $\chi$ by reversing the above construction, which will give particular derivatives of the $\delta$ distribution contracted with constant tensors for $f^{\mu\nu}$ and $\chi$. This statement is equivalent to saying that the graviton and inflaton perturbations are only locally observable in the particular differentiated form in which they appear in the linearised Weyl tensor and in $C_{\mu\nu}$, but not directly. In particular, the linearised Ricci scalar of constant-field hypersurfaces $\mathcal{R}$, given by equation~\eqref{appendix_geometric_ricci3}, appears as the spatial trace of $C_{\mu\nu}$:
\begin{equation}
\label{spatial_trace_cmunu}
\delta^{ij} C_{ij} = \frac{1}{n-2} \left[ \partial_i \partial_j h^{ij} - \laplace \left( h + h_{00} \right) \right] + 2 \laplace \left( \frac{H a}{\phi'} \phi^{(1)} \right) = \frac{1}{n-2} a^2 \mathcal{R}^{(1)} \eqend{.}
\end{equation}

\section{Local observables in terms of tensor and scalar perturbations}
\label{correlators}

In the previous section we expressed our compactly supported observable $I(f,\chi)$ in terms of the local observables $C^{(1)}_{\alpha\mu\beta\nu}$ and $C_{\mu\nu}$. In this section we express these local observables in terms of the tensor perturbation and Sasaki--Mukhanov variable, which describes the scalar perturbation. We follow the notation of ref.~\cite{froeb_jcap_2014}.

The metric and scalar-field perturbations that allow Fourier decomposition can be given as
\begin{equations}[h_decomp]
h_{00} &= S + 2 X_0' + 2 H a X_0 \eqend{,} \\
h_{0k} &= V_k + X_k' + \partial_k X_0 \eqend{,} \\
h_{kl} &= H_{kl} + \delta_{kl} \Sigma + 2 \partial_{(k} X_{l)} - 2 H a \delta_{kl} X_0 \eqend{,} \\
\phi^{(1)} &= \Psi - X_0 \phi' \eqend{.}
\end{equations}
The vector perturbation $V_k$ is transverse and the tensor perturbation $H_{kl}$ is transverse and traceless. The gauge transformation~\eqref{gauge_trafo_perts} transforms $X_\mu$ as $\delta X_\mu = \xi_\mu$. All other variables on the right-hand side of equations~\eqref{h_decomp} are gauge-invariant, even though they are in general non-local functions of $h_{\mu\nu}$ and $\phi^{(1)}$, unlike $C^{(1)}_{\alpha\mu\beta\nu}$ and $C_{\mu\nu}$. It is convenient to introduce the Sasaki--Mukhanov variable $Q$ as follows~\cite{mukhanov_et_al_pr_1992}:
\begin{equation}
Q \equiv \frac{2 H a}{\phi'} \Psi - \Sigma \eqend{,}
\end{equation}
so that
\begin{equation}
\label{phi_decomp}
\phi^{(1)} = \frac{\phi'}{2 H a} \left( \Sigma + Q - 2 H a X_0 \right) \eqend{.}
\end{equation}
The gauge-invariant variables in \eqref{h_decomp} satisfy the following field equations~\cite{froeb_jcap_2014}, which result from $E^{(1)}_{\mu\nu} = 0$ and $F^{(1)} = 0$:
\begin{equations}[eom]
\partial^2 H_{kl} &= (n-2) H a H_{kl}' \eqend{,} \\
V_k &= 0 \eqend{,} \\
\laplace \Sigma &= - H a \epsilon Q' \eqend{,} \\
\laplace S &= - (n-3) H a \epsilon Q' \eqend{,} \\
\partial^2 Q &= (n-2+2\delta) H a Q' \eqend{,}
\end{equations}
where $\laplace \equiv \delta^{kl} \partial_k \partial_l$ as before. On-shell, the linearised Weyl tensor $C^{(1)}_{\alpha\mu\beta\nu}$ is expressed in terms of the tensor perturbation $H_{kl}$ and the Sasaki--Mukhanov variable $Q$ as~\cite{froeb_jcap_2014}
\begin{equations}[weyl_expressions]
C^{(1)}_{0j0l} &= \frac{a^2}{2} \left[ (n-3) H a H_{jl}' - \laplace H_{jl} \right] - \frac{n-3}{2 (n-1)} H a^3 \epsilon \Pi_{jl} Q' \eqend{,} \\
C^{(1)}_{0jkl} &= - a^2 \partial_{[k} H'_{l]j} \eqend{,} \\
C^{(1)}_{ijkl} &= a^2 \left[ - \partial_i \partial_{[k} H_{l]j} + \partial_j \partial_{[k} H_{l]i} + H a \left( \eta_{i[k} H'_{l]j} - \eta_{j[k} H'_{l]i} \right) \right] - \frac{H a^3 \epsilon}{n-1} \left( \Pi_{i[k} \eta_{l]j} - \Pi_{j[k} \eta_{l]i} \right) Q' \eqend{,}
\end{equations}
where $\Pi_{kl}$ is the traceless projection operator
\begin{equation}
\Pi_{kl} = \delta_{kl} - (n-1) \frac{\partial_k \partial_l}{\laplace} \eqend{.}
\end{equation}
Notice that the variables $H_{kl}$ and $Q$ are differentiated either once with respect to $\eta$ or twice with respect to space coordinates in equations~\eqref{weyl_expressions}.

Now we find an expression analogous to~\eqref{weyl_expressions} for $C_{\mu\nu}$.  Substituting the expansion of $h_{\mu\nu}$~\eqref{h_decomp} and $\phi^{(1)}$~\eqref{phi_decomp} into equation~\eqref{C_expression}, we find
\begin{equations}[cmunu_decomp]
C_{00} &= Q'' - \frac{1}{n-1} \laplace \left( S + \Sigma \right) \eqend{,} \\
C_{0k} &= \partial_k Q' - \frac{1}{n-2} \laplace V_k \eqend{,} \\
C_{kl} &= - \frac{1}{n-2} \left( \partial^2 H_{kl} + 2 \partial_{(k} V_{l)}' \right) + \partial_k \partial_l Q - \frac{1}{(n-1) (n-2)} \Pi_{kl} \laplace \left( S + \Sigma \right) \eqend{.}
\end{equations}
Using the equations of motion~\eqref{eom} to simplify these expressions on-shell we obtain
\begin{equations}[cmunu_decomp_onshell]
C_{00} &= \laplace Q - \left( n-2 + 2 \delta - \frac{n-2}{n-1} \epsilon \right) H a Q' \eqend{,} \\
C_{0k} &= \partial_k Q' \eqend{,} \\
C_{kl} &= - H a H_{kl}' + \partial_k \partial_l Q + \frac{1}{n-1} \Pi_{kl} H a \epsilon Q' \eqend{,}
\end{equations}
and we recover the well-known on-shell expression for the linearised Ricci scalar of constant-field hypersurfaces~\eqref{spatial_trace_cmunu}
\begin{equation}
\label{linearised_3ricci_c_q}
\mathcal{R}^{(1)} = (n-2) a^{-2} \delta^{ij} C_{ij} = (n-2) a^{-2} \laplace Q \eqend{.}
\end{equation}
Notice that the variables $H_{kl}$ and $Q$ are again differentiated either once with respect to $\eta$ or twice with respect to space coordinates, as in the case of the linearised Weyl tensor~\eqref{weyl_expressions}. This makes the IR behaviour of the correlators of $C^{(1)}_{\alpha\mu\beta\nu}$ and $C_{\mu\nu}$, and hence of $I(f,\chi)$, upon quantisation much better than that for the variables $H_{kl}$ or $Q$, as we shall see in the next section.

\section{Infrared behaviour of the compactly supported observable}
\label{ir_behaviour}

To discuss the IR behaviour of the correlators of our compactly supported (quantum) observable $I(f,\chi)$, or, equivalently, that of the observables $C^{(1)}_{\alpha\mu\beta\nu}$ and $C_{\mu\nu}$, we first review the quantisation of single-field inflation at linear order following~\cite{froeb_jcap_2014}.

We first write the field equations for $H_{kl}$ and $Q$ listed in equation~\eqref{eom} as
\begin{equations}[eom_for_H_and_Q]
&H_{kl}'' + (n-2) H a H_{kl}' - \laplace H_{kl} = 0 \eqend{,} \\
&Q'' + (n-2 + 2\delta) H a Q' - \laplace Q = 0 \eqend{.} \label{eom_for_H_and_Q_no_2}
\end{equations}
We then expand these fields as
\begin{equations}[H_Q_mode_expansion]
H_{kl}(\eta,\vec{x}) &= \int \sum_{\sigma} e^{(\sigma)}_{kl}(\vec{p}) a_{(\vec{p},\sigma)} f_p(\eta) \mathe^{\mathi \vec{p} \vec{x}} \frac{\total^{n-1} p}
{(2\pi)^{n-1}} + \text{h.c.} \eqend{,} \\
Q(\eta,\vec{x}) &= \int b_{\vec{p}} q_p(\eta) \mathe^{\mathi \vec{p} \vec{x}} \frac{\total^{n-1} p}{(2\pi)^{n-1}} + \text{h.c.} \eqend{,}
\end{equations}
where $e^{(\sigma)}_{kl}(\vec{p})$ are the polarisation tensors satisfying $p^k e_{kl}^{(\sigma)}(\vec{p}) = \eta^{kl} e_{kl}^{(\sigma)}(\vec{p}) = 0$ and the orthonormality condition $e^{(\sigma)}_{kl}(\vec{p}) e^{(\sigma')*}_{kl}(\vec{p}) = 2 \delta^{\sigma\sigma'}$. We have assumed rotational invariance so that $f_p(\eta)$ and $q_p(\eta)$ depend only on $p = \abs{\vec{p}}$ and not on the direction of $\vec{p}$. For the normalised commutation relations $[ a_{(\vec{p},\sigma)}, a^\dagger_{(\vec{p}',\sigma')} ] = (2\pi)^{n-1} \delta_{\sigma\sigma'} \delta^{n-1}(\vec{p}-\vec{p}')$ and $[ b_{\vec{p}}, b^\dagger_{\vec{p}'} ] = (2\pi)^{n-1} \delta^{n-1}(\vec{p}-\vec{p}')$ to hold, the functions $f_p$ and $q_p$ need to satisfy the normalisation conditions~\cite{froeb_jcap_2014}
\begin{equations}[f_q_normalisation]
f_p(\eta) f_p^{*\prime}(\eta) - f'_p(\eta) f^*_p(\eta) &= \mathi \kappa^2 a^{2-n} \eqend{,} \\
q_p(\eta) q_p^{*\prime}(\eta) - q'_p(\eta) q^*_p(\eta) &= \frac{2 \mathi \kappa^2 a^{2-n}}{(n-2) \epsilon} \eqend{.}
\end{equations}
The vacuum state $\ket{0}$ is defined by $a_{\vec{p}} \ket{0} = b_{\vec{p}} \ket{0} = 0$ for all $\vec{p}$. Then the correlation function for the tensor perturbation is given by
\begin{splitequation}
\label{hkl_corr}
\Delta^{H}_{ijkl}(\eta,\eta',\abs{\vec{x}-\vec{x}'}) &\equiv \bra{0} H_{ij}(\eta,\vec{x}) H_{kl}(\eta',\vec{x}') \ket{0} \\
&= \int \left( 2 P_{k(i} P_{j)l} - \frac{2}{n-2} P_{ij} P_{kl} \right) f_p(\eta) f^*_p(\eta') \mathe^{\mathi \vec{p} (\vec{x}-\vec{x}')} \frac{\total^{n-1} p}{(2\pi)^{n-1}} \eqend{,}
\end{splitequation}
where
\begin{equation}
P_{kl} \equiv \delta_{kl} - \frac{\partial_k \partial_l}{\laplace} \eqend{,}
\end{equation}
and the correlation function for the Sasaki--Mukhanov variable reads
\begin{splitequation}
\label{q_corr}
\Delta^{Q}(\eta,\eta',\abs{\vec{x}-\vec{x}'}) &\equiv \bra{0} Q(\eta, \vec{x}) Q(\eta', \vec{x}') \ket{0} \\
&= \int q_p(\eta) q^*_p(\eta') \mathe^{\mathi \vec{p} (\vec{x}-\vec{x}')} \frac{\total^{n-1} p}{(2\pi)^{n-1}} \eqend{.}
\end{splitequation}
These correlators are IR divergent, \ie, the $p$-integral diverges in the IR in many applications, most notably in the slow-roll approximation in single-field inflation. That is, $f_p(\eta)$ and $q_p(\eta)$ diverge like $p^{-(n-1)/2}$ or stronger as $p \to 0$ in many applications, and thus the above Fourier integrals are not convergent unless one introduces an IR cutoff.

For the compactly supported observable $I(f,\chi)$, or equivalently for $C^{(1)}_{\alpha\mu\beta\nu}$ and $C_{\mu\nu}$, the fields $H_{kl}$ and $Q$ occur with one time derivative or two space derivatives, as we observed before. This implies that the IR divergence of the correlation functions are reduced at least by a factor of $p^4$ as we now show. From the equations of motion~\eqref{eom_for_H_and_Q} and the mode expansion~\eqref{H_Q_mode_expansion}, it follows that the functions $f_p$ and $q_p$ satisfy
\begin{equations}
\left[ a^{n-2}(\eta) f'_p(\eta) \right]' &= - p^2 a^{n-2}(\eta) f_p(\eta) \eqend{,} \\
\left[ \epsilon(\eta) a^{n-2}(\eta) q'_p(\eta) \right]' &= - p^2 \epsilon(\eta) a^{n-2}(\eta) q_p(\eta) \eqend{.}
\end{equations}
We can solve these equations for small $p^2$, where we can neglect the right-hand side in comparison with the left-hand one because of the explicit factors of $p^2$. This gives
\begin{equations}[ABCD]
q_p(\eta) &= \mathi A(p) \left[ 1 + \bigo{p^2} \right] + B(p) \left[ \int \frac{\total \eta}{\epsilon(\eta) a^{n-2}(\eta)} + \bigo{p^2} \right] \eqend{,}\\
f_p(\eta) &= \mathi C(p) \left[ 1 + \bigo{p^2} \right] + D(p) \left[ \int \frac{\total \eta}{a^{n-2}(\eta)} + \bigo{p^2} \right] \eqend{,}
\end{equations}
where without loss of generality we choose $A(p)$ and $C(p)$ to be real and positive, since an overall phase factor
affects neither the normalisation conditions~\eqref{f_q_normalisation} nor the correlation functions~\eqref{hkl_corr} and~\eqref{q_corr}. The normalisation conditions then imply
\begin{equations}
A(p) \left[ B(p) + B^*(p) \right] &= \frac{2\kappa^2}{n-2} + \bigo{p^2} \eqend{,} \\
C(p) \left[ D(p) + D^*(p) \right] &= \kappa^2 + \bigo{p^2} \eqend{.}
\end{equations}
The choice of the quantum state determines $A(p)$ and $C(p)$ subject to this condition. For the natural quantum state in many interesting applications, in particular for the Bunch--Davies vacuum in slow-roll single-field inflationary models~\cite{mukhanov_textbook,dodelson_textbook,baumann_lectures_2009}, $A(p)$ and $C(p)$ diverge like $p^{-(n-1)/2}$ or faster, leading to IR divergence of the correlators for $Q$ and $H_{kl}$ as explained above. This implies that the real parts of $B(p)$ and $D(p)$ tend to zero in this limit, and if their imaginary parts also tend to zero (as happens for slow-roll inflationary models), they do not cause any IR divergence. Therefore, the leading IR-divergent terms in the limit $p \to 0$ are given by $A(p)$ and $C(p)$, which are annihilated by a derivative with respect to $\eta$, and thus the leading terms in the $p \to 0$ limit in $q_p'(\eta)$ and $f_p'(\eta)$ are suppressed by a factor of $p^2$ as compared to $q_p(\eta)$ and $f_p(\eta)$ themselves. The leading behaviour of the second-order space derivatives of the mode functions $q_p(\eta) \mathe^{\mathi\vec{p}\vec{x}}$ and $f_p(\eta) \mathe^{\mathi\vec{p}\vec{x}}$ are obviously also suppressed by a factor of $p^2$. Since the variables $Q$ and $H_{kl}$ are differentiated once with respect to $\eta$ or twice with respect to $\vec{x}$ in the expression for $C^{(1)}_{\alpha\mu\beta\nu}$ and $C_{\mu\nu}$, the leading behaviour of their mode functions in the limit $p \to 0$ is thus suppressed by $p^2$. This implies that the integrand of the $p$-integration in the correlators for quantum operators $C^{(1)}_{\alpha\mu\beta\nu}$ and $C_{\mu\nu}$, and hence that for $I(f,\chi)$, are suppressed by $p^4$. This suppression is sufficient to make the correlator for $I(f,\chi)$ IR finite in most applications relevant to the physics of the early universe.

To make these general considerations more accessible, let us study an example. Since at tree level there are no UV divergences, we set $n = 4$ in the following. In the slow-roll approximation of single-field inflation, one has
\begin{equations}[slow_roll_data]
a(\eta) & = a_0 (-\eta)^{- \frac{1}{1-\epsilon}} \eqend{,} \\
H(\eta) & = \frac{1}{(1-\epsilon) a_0} (-\eta)^{\frac{\epsilon}{1-\epsilon}} \eqend{,} \\
\epsilon(\eta) & = \epsilon_0 (-\eta)^{-2\delta}\eqend{,}
\end{equations}
where $a_0$ and $\epsilon_0$ are positive constants,
to first order in $\epsilon$ and $\delta$. We note that these relations are exact for constant positive $\epsilon$
(and, consequently, with $\delta=0$) even if it is large.  For the mode functions, we then have~\cite{froeb_jcap_2014}
\begin{equations}[mode_functions]
f_p(\eta) &= \frac{\sqrt{\pi}}{2} \kappa (1-\epsilon) H(\eta) (-\eta)^\frac{3}{2} \hankel1_{\frac{3}{2} + \frac{\epsilon}{1-\epsilon}}\left( -p \eta \right) 
\eqend{,} \\
q_p(\eta) &= \frac{\sqrt{\pi}}{2} \kappa (1-\epsilon) \frac{H(\eta)}{\sqrt{\epsilon(\eta)}} (-\eta)^\frac{3}{2} \hankel1_{\frac{3}{2} + \frac{\epsilon}{1-\epsilon} + \delta}\left( - p \eta \right) \eqend{,}
\end{equations}
again to first order in $\epsilon$ and $\delta$. (These expressions and the results that follow are exact for constant $\epsilon < 1$.) The small $p$ approximation of these functions can be found by using
\begin{equations}
\hankel1_\nu(z) &= \frac{\mathrm{J}_{-\nu}(z) - \mathe^{-\nu\pi\mathi} \mathrm{J}_\nu(z)}{\mathi \sin \nu \pi} \eqend{,} \\
\mathrm{J}_\nu(z) &= \frac{1}{\Gamma(1+\nu)} \left( \frac{z}{2} \right)^\nu \left[ 1 + \bigo{z^2} \right] \eqend{.}
\end{equations}
Thus, the coefficients in equation~\eqref{ABCD} in these models are given by 
\begin{equations}
A(p) & = - \frac{\kappa \Gamma\left( \frac{3}{2} + \frac{\epsilon}{1-\epsilon}+\delta \right)}{2 \sqrt{\pi \epsilon_0} \, a_0} p^{- \frac{3}{2} - \frac{\epsilon}{1-\epsilon} - \delta} \eqend{,} \label{slow_roll_coeff_a} \\
B(p) & = - \kappa \left[ 1 - \mathi \pi \left( \frac{\epsilon}{1-\epsilon} + \delta \right) \right] \frac{\sqrt{\pi \epsilon_0} \, a_0}{\Gamma\left( \frac{3}{2} + \frac{\epsilon}{1-\epsilon}+\delta \right)} p^{\frac{3}{2} + \frac{\epsilon}{1-\epsilon} + \delta} \eqend{,} \label{slow_roll_coeff_b} \\
C(p) & = - \frac{\kappa \Gamma\left( \frac{3}{2} + \frac{\epsilon}{1-\epsilon} \right)}{2 \sqrt{\pi} \, a_0} p^{- \frac{3}{2} - \frac{\epsilon}{1-\epsilon}} \eqend{,} \label{slow_roll_coeff_c} \\
D(p) & = - \kappa \left( 1 - \mathi \pi \frac{\epsilon}{1-\epsilon} \right) \frac{\sqrt{\pi} \, a_0}{\Gamma\left( \frac{3}{2} + \frac{\epsilon}{1-\epsilon} \right)} p^{\frac{3}{2} + \frac{\epsilon}{1-\epsilon}}
\eqend{.} \label{slow_roll_coeff_d}
\end{equations}
We note that these coefficients satisfy the properties mentioned before, with $A(p)$ and $C(p)$ diverging as $p \to 0$ and $B(p)$ and $D(p)$ vanishing in this limit. Thus, they lead to IR divergences of the correlators for $Q$ and $H_{kl}$ (as long as $\epsilon + \delta \geq 0$ and $\epsilon \geq 0$, respectively), and one time derivative or two space derivatives produce an extra factor of $p^4$ in the integrand for the $p$-integration for the correlators, rendering them IR finite (as long as $\frac{\epsilon}{1-\epsilon} + \delta < 2$ and $\epsilon < \frac{2}{3}$, respectively). 
The IR-finiteness condition is always satisfied in slow-roll inflation because we have $\epsilon, \delta \ll 1$. 

For constant $\epsilon$ the correlators for $H_{kl}$ and $Q$ are IR divergent if $\abs{ \frac{3}{2} + \frac{\epsilon}{1-\epsilon} } \geq \frac{3}{2}$, \ie, if $0 \leq \epsilon \leq \frac{3}{2}$, which includes the matter-dominated universe, $\epsilon = \frac{3}{2}$. On the other hand the correlators for $I(f,\chi)$ are IR divergent only for $\abs{ \frac{3}{2} + \frac{\epsilon}{1-\epsilon} } \geq \frac{7}{2}$, \ie, for $\frac{2}{3} \leq \epsilon \leq \frac{5}{4}$. Thus, the correlators of $I(f,\chi)$ are IR finite for the matter ($\epsilon = \frac{3}{2}$) and radiation ($\epsilon = 2$) dominated universes. Even for $\frac{2}{3} \leq \epsilon \leq \frac{5}{4}$, the conclusion of IR divergence assumes a choice of the mode functions, and hence of the vacuum state, which respects the rescaling symmetry of the field equation $\eta \to \lambda \eta, \vec{x} \to \lambda \vec{x}$ (which we discuss below). Although this choice is generally accepted as the right choice for slow-roll inflation, it may not be the right choice for these values of $\epsilon$. Furthermore, it has been shown~\cite{ford_parker_prd_1977b,janssen_prokopec_prd_2009} that IR-regular quantum states result if the state during the preceding spacetime evolution did not suffer from IR divergences.\footnote{This is a consequence of the mathematical fact that given a correlation function which is a well-defined distribution in some time interval (in particular without IR divergences), and which satisfies a linear hyperbolic equation (the equation of motion), it is a well-defined distribution for all times, see, \eg,~\cite{baerbook}.} That is, if a universe with $\epsilon$ satisfying $\frac{2}{3} \leq \epsilon \leq \frac{5}{4}$ was preceded by a less IR-singular era (such as in the standard Big Bang cosmological model with an inflationary phase~\cite{mukhanov_textbook,dodelson_textbook,baumann_lectures_2009}), the quantum state will \emph{not} be the Bunch--Davies-type vacuum with the mode functions~\eqref{mode_functions}, but an excited state which is less singular in the IR.

Let us now turn to the closely related problems of secular growth, \ie, divergences in the late-time limit $\eta \to 0$, $a(\eta) \to \infty$, and the growth of the correlators for large spatial separations. To leading order in $\epsilon$ and $\delta$, because of IR divergences the correlators for $H_{kl}$ and $Q$ grow logarithmically for $r \equiv \abs{\vec{x}-\vec{x}'} \to \infty$ and equal time ($\eta = \eta'$) as
\begin{equations}[IR_dS_limit]
\Delta^{H}_{ijkl}(\eta,\eta,r) &\approx - \frac{\kappa^2 H^2}{10 \pi^2} \left( \delta_{ik} \delta_{jl} + \delta_{il} \delta_{jk} - \frac{2}{3} \delta_{ij} \delta_{kl} \right) \ln\left( \xi r \right) \eqend{,} \\
\Delta^{Q}(\eta,\eta,r) &\approx - \frac{\kappa^2 H^2}{4 \pi^2 \epsilon} \ln\left( \xi r \right) \eqend{,}
\end{equations}
where $\xi$ is the IR cutoff in the momentum integral. Furthermore, if the \emph{physical} distance $a(\eta) r$ between the two points is fixed, then they increase like $\ln\left[ a(\eta) \right]$ as the universe expands (the secular growth) because $\ln r = \ln\left[ a(\eta) r \right] - \ln\left[ a(\eta) \right]$. The long-distance and secular growth of the correlators appears to cause physically observable quantities to have this growth behaviour, which has been debated in the past several years.

However, as we have seen, the integrands for the $p$-integration in the correlators for the variables $C^{(1)}_{\alpha\mu\beta\nu}$ and $C_{\mu\nu}$ have an extra factor of $p^4$ compared to the integrand for the correlators for $Q$ and $H_{kl}$. For small $p$, the behaviour of the integrand of the $Q$ correlator~\eqref{q_corr} is dominated by the coefficient $A(p)$~\eqref{slow_roll_coeff_a} in the mode functions $q_p(\eta)$, which leads to a behaviour of the $p$-integral of the form $\int p^{-1-\frac{2\epsilon}{1-\epsilon}-2\delta} \total p$. Similarly, the 
$p$-integration of the $H_{kl}$ correlator~\eqref{hkl_corr} is dominated by the coefficient $C(p)$~\eqref{slow_roll_coeff_c} in the mode functions $f_p(\eta)$, and behaves like $\int p^{-1-\frac{2\epsilon}{1-\epsilon}} \total p$. Thus, the integrand for the $p$-integration in the correlators of $C^{(1)}_{\alpha\mu\beta\nu}$ and $C_{\mu\nu}$ behaves like either $\int p^{3-\frac{2\epsilon}{1-\epsilon}-2\delta} \total p$ or $\int p^{3-\frac{2\epsilon}{1-\epsilon}} \total p$. By power counting we conclude that the leading behaviour of these correlators as $r \to \infty$ is either $r^{-4+\frac{2\epsilon}{1-\epsilon}+2\delta}$ or $r^{-4+\frac{2\epsilon}{1-\epsilon}}$. Thus, it is clear that there is no long-distance growth in the correlators for $C^{(1)}_{\alpha\mu\beta\nu}$ and $C_{\mu\nu}$, and thus for the compactly supported observable $I(f, \chi)$ for small $\epsilon$ and $\delta$, and for $0 \leq \epsilon < \frac{2}{3}$ or $\epsilon > \frac{5}{4}$ with $\delta = 0$.

The lack of secular growth for these correlators can be shown using a simple scaling argument. Namely, to first order in the slow-roll parameters $\epsilon$ and $\delta$ (or for constant, but arbitrary $\epsilon$), we find from equations~\eqref{slow_roll_data} that
\begin{equations}[rescalings]
a(\lambda \eta) &= \lambda^{- \frac{1}{1-\epsilon}} a(\eta) \eqend{,} \\
[1-\epsilon(\lambda \eta)] H(\lambda \eta) &= \lambda^\frac{\epsilon}{1-\epsilon} [1-\epsilon(\eta)] H(\eta) \eqend{,} \\
\epsilon(\lambda \eta) &= \lambda^{-2\delta} \epsilon(\eta) \eqend{,}
\end{equations}
where $\lambda$ is a positive constant, and thus for the mode functions~\eqref{mode_functions}
\begin{equations}
f_{p/\lambda}(\lambda \eta) &= \lambda^{\frac{3}{2} + \frac{\epsilon}{1-\epsilon}} f_p(\eta) \eqend{,} \\
q_{p/\lambda}(\lambda \eta) &= \lambda^{\frac{3}{2} + \frac{\epsilon}{1-\epsilon} +  \delta} q_p(\eta) \eqend{.}
\end{equations}
By substituting this scaling behaviour into the correlation functions~\eqref{q_corr} and~\eqref{hkl_corr}, we thus find the following na{\"\i}ve scaling properties (with $r \equiv \abs{\vec{x}-\vec{x}'}$):
\begin{equations}[scaling_correlators]
\Delta^Q(\lambda \eta, \lambda \eta', \lambda r) &= \lambda^{\frac{2 \epsilon}{1-\epsilon} + 2 \delta} \Delta^Q(\eta, \eta', r) \eqend{,}
\label{scaling_Q} \\
\Delta^H_{ijkl}(\lambda \eta, \lambda \eta', \lambda r) &= \lambda^\frac{2 \epsilon}{1-\epsilon} \Delta^H_{ijkl} (\eta, \eta', r) \eqend{.}
\end{equations}
However, the IR cutoff $\xi$ that one needs to introduce breaks these scaling properties, since the scaling $p \to p/\lambda$ is not accompanied by a corresponding scaling in $\xi$. Thus, the scaling~\eqref{scaling_correlators} will be valid whenever the IR cutoff can be removed. For definiteness, we consider the correlator for $Q$ with two derivatives at each point, all in the $i$-th direction. The scaling law~\eqref{scaling_Q} can be used to find
\begin{equation}
\label{scaling_law}
a^{-4}(\lambda \eta) \partial^2_{\lambda x^i} \partial^2_{\lambda x^{\prime i}} \Delta^Q(\lambda \eta, \lambda \eta', \lambda r) = \lambda^{\frac{6 \epsilon}{1-\epsilon} + 2 \delta} a^{-4}(\eta) \partial_{x^i}^2 \partial_{x^{\prime i}}^2 \Delta^Q(\eta, \eta', r) \eqend{.}
\end{equation}
In de~Sitter space, where $\epsilon = \delta = 0$, this shows that the correlator is unchanged under the rescaling, while in the general case there is only a mild dependence on $\lambda$. However, to properly account for secular growth in the late-time limit we should not fix the coordinate distance $r$, but the physical distance $r_\text{phys}$ between the two points, defined as
\begin{equation}
r_\text{phys} \equiv a(\eta) r \eqend{,}
\end{equation}
with $\eta' = \eta$.  This motivates us to define
\begin{equation}
\hat{\Delta}^Q_\text{phys}\left( \eta, r_\text{phys} \right) \equiv \partial_{x^i_\text{phys}}^2 \partial_{x^{\prime i}_\textrm{phys}}^2 \Delta^Q\left( \eta, \eta, a^{-1}(\eta) r_\text{phys} \right) \eqend{,}
\end{equation}
where $x^i_\text{phys} \equiv a(\eta) x^i$ and $x^{\prime i}_\text{phys} \equiv a(\eta) x^{\prime i}$. Then the scaling law~\eqref{scaling_law} leads to
\begin{equation}
\hat{\Delta}_\text{phys}^Q \left( \lambda\eta, r_\text{phys} \right) = \lambda^{\frac{6\epsilon}{1-\epsilon}+2\delta} \hat{\Delta}_\text{phys}^Q\left( \eta, \lambda^{\frac{\epsilon}{1-\epsilon}} r_\text{phys} \right) \eqend{.}
\end{equation}
Let us specialise to the case $0 \leq \epsilon < \frac{2}{3}$. The late-time limit $\eta \to 0$ then corresponds to taking $\lambda \to 0$ in this expression. Again, in the de~Sitter limit $\epsilon = \delta = 0$, the correlator is unchanged also under this rescaling that holds $r_\text{phys}$ fixed, and thus we conclude that there is no secular growth at late times. In the general case, there is mild dependence on $\lambda$.  We find that this correlator tends to a constant if $\delta = 0$ but to zero if $\delta > 0$ in the limit $\lambda \to 0$. The same conclusions can be drawn in exactly the same way for the normalised time derivative of $\Delta^Q$, given by $a(\eta)^{-1} \partial_\eta \Delta^Q$, and all normalised derivatives of $\Delta^H_{ijkl}$. Considering then the correlators of ${C^{(1)\alpha\mu}}_{\beta\nu}$ and ${C^\mu}_\nu$ (where the indices are raised with the background metric $\tilde{g}^{\mu\nu} = a^{-2} \eta^{\mu\nu}$), we see from the explicit expressions~\eqref{weyl_expressions} and~\eqref{cmunu_decomp_onshell} that all derivatives are normalised, possibly up to factors of $H$ or $\epsilon$ which only show a mild dependence on $\lambda$ upon rescaling~\eqref{rescalings}. This implies that there is only mild secular growth, if any, predicted by the scaling argument for the correlators of ${C^{(1)\alpha\mu}}_{\beta\nu}$ and ${C^\mu}_\nu$, and thus for the correlators of our gauge-invariant observable $I(f,\chi)$.

\section{Relation of the compactly supported observables to the power spectra}

\label{section:power_spectra}

To recover from our observables the well-known tensor and scalar power spectra~\cite{mukhanov_textbook,dodelson_textbook,baumann_lectures_2009}
\begin{equations}[power_spectra]
\mathcal{P}_\text{H}^2(\abs{\vec{k}},\eta) &= \frac{\abs{\vec{k}}^3}{2 \pi^2} \delta^{ik} \delta^{jl} \int \bra{0} H_{ij}(\vec{x},\eta) H_{kl}(\vec{0},\eta) \ket{0} \mathe^{- \mathi \vec{k} \vec{x}} \total^3 x \approx \frac{\kappa^2 H^2}{\pi^2} \eqend{,} \\
\mathcal{P}_\text{S}^2(\abs{\vec{k}},\eta) &= \frac{\abs{\vec{k}}^3}{8 \pi^2} \int \bra{0} Q(\vec{x},\eta) Q(\vec{0},\eta) \ket{0} \mathe^{- \mathi \vec{k} \vec{x}} \total^3 x \approx \frac{\kappa^2 H^2}{16 \pi^2 \epsilon} \eqend{,}
\end{equations}
where the last expressions are the tree-level results for the Bunch--Davies vacuum at leading order in slow-roll, we follow the steps presented in ref.~\cite{froeb_jcap_2014} for the tensor spectrum. First we express the tensor perturbation and the Sasaki--Mukhanov variable in terms of the linearised Weyl tensor and the spatial trace of the tensor $C_{\mu\nu}$ [see equation~\eqref{linearised_3ricci_c_q}] according to\footnote{Note that the expression for $H_{kl}$ is different from the one given in ref.~\cite{froeb_jcap_2014}. Both agree on-shell, but the one presented here has the advantage of not requiring time derivatives.}
\begin{equations}[hmunu_q_in_weyl_cmunu]
\laplace^3 H_{kl} &= - a^{-2} P_{kl}^{\mu\nu\rho\sigma} C^{(1)}_{\mu\nu\rho\sigma} \eqend{,} \\
P_{kl}^{\mu\nu\rho\sigma} &= 2 \delta^\mu_i \delta^\nu_k \delta^\rho_j \delta^\sigma_l \laplace \partial^i \partial^j - 2 \delta^\mu_0 \delta^\nu_i \delta^\rho_0 \delta^\sigma_j \left( \laplace \delta_{kl} - \partial_k \partial_l \right) \partial^i \partial^j + 2 H a \delta^\mu_0 \delta^\rho_i \delta^\nu_{(k} \delta^\sigma_{l)} \laplace \partial^i \eqend{,} \\
\laplace Q &= \delta^{ij} C_{ij} \eqend{,}
\end{equations}
and then rewrite the power spectra~\eqref{power_spectra} in terms of the right-hand sides (using the symmetries of the Weyl tensor):
\begin{equations}[power_spectra_weyl_cmunu]
\begin{split}
\mathcal{P}_\text{H}^2(\abs{\vec{k}},\eta) &= \frac{1}{2 \pi^2 \abs{\vec{k}}^9} \delta^{ik} \delta^{jl} a^{-4} \int \bra{0} P_{ij}^{\alpha\beta\gamma\delta} C^{(1)}_{\alpha\beta\gamma\delta}(\vec{x},\eta) P_{kl}^{\mu\nu\rho\sigma} C^{(1)}_{\mu\nu\rho\sigma}(\vec{0},\eta) \ket{0} \mathe^{- \mathi \vec{k} \vec{x}} \total^3 x \\
&= \frac{1}{\pi^2 \abs{\vec{k}}^5} a^{-4} \left[ 2 \bar{\eta}^{\alpha\mu} \bar{\eta}^{\gamma\rho} \vec{k}^\beta \vec{k}^\delta \vec{k}^\nu \vec{k}^\sigma - 4 \mathi H a \bar{\eta}^{\alpha\mu} \bar{\eta}^{\gamma\rho} \delta^\nu_0 \vec{k}^\beta \vec{k}^\delta \vec{k}^\sigma + H^2 a^2 \delta^\alpha_0 \delta^\mu_0 \bar{\eta}^{\beta(\nu} \bar{\eta}^{\sigma)\delta} \vec{k}^\gamma \vec{k}^\rho \right] \\
&\qquad\times \int \bra{0} C^{(1)}_{\alpha\beta\gamma\delta}(\vec{x},\eta) C^{(1)}_{\mu\nu\rho\sigma}(\vec{0},\eta) \ket{0} \mathe^{- \mathi \vec{k} \vec{x}} \total^3 x \eqend{,}
\end{split} \\
\mathcal{P}_\text{S}^2(\abs{\vec{k}},\eta) &= \frac{1}{8 \pi^2 \abs{\vec{k}}} \delta^{ij} \delta^{kl} \int \bra{0} C_{ij}(\vec{x},\eta) C_{kl}(\vec{0},\eta) \ket{0} \mathe^{- \mathi \vec{k} \vec{x}} \total^3 x \eqend{,}
\end{equations}
where we defined $\bar{\eta}^{\mu\nu} \equiv \eta^{\mu\nu} + \delta_0^\mu \delta_0^\nu$. Since the relations~\eqref{hmunu_q_in_weyl_cmunu} are (on-shell) identities, this gives exactly the same power spectra. This shows clearly two crucial points: a) the power spectra are gauge-invariant observables (at linear order), since they are obtained from the manifestly gauge invariant linearised Weyl tensor and $C_{\mu\nu}$, and b) they are non-local (because of the integration over the full three-dimensional space) and possibly IR divergent as $\abs{\vec{k}} \to 0$ because of the explicit inverse powers of $\abs{\vec{k}}$.

To make contact with our observable $I(f,\chi)$, we use our result in the form~\eqref{final_equivalence}
\begin{equation}
I(f,\chi) = - \frac{1}{2} \int \hat{w}^{\alpha\mu\beta\nu} C_{\alpha\mu\beta\nu}^{(1)} \sqrt{-g} \total^4 x - \int \hat{u}^{\mu\nu} C_{\mu\nu} \sqrt{-g} \total^4 x \eqend{,}
\end{equation}
and take the smearing functions $\hat{w}^{\alpha\mu\beta\nu}$ and $\hat{u}^{\mu\nu}$ such that the correlators of equation~\eqref{power_spectra_weyl_cmunu} emerge. That is, for the scalar power spectrum we take
\begin{equation}
\label{u_testfunction_example}
\hat{u}^{\mu\nu}(\vec{x},\tau) = \frac{1}{\sqrt{8 \pi^2 \abs{\vec{k}}}} \bar{\eta}^{\mu\nu} a(\eta)^{-n} f^\delta_{\eta}(\tau) \mathe^{- \mathi \vec{k} \vec{x}} g_{\delta'}(\vec{x}) \eqend{,}
\end{equation}
where $f^\delta_\eta(\tau)$ is a real smooth function of compact support depending on a parameter $\delta$ such that
\begin{equation}
\lim_{\delta \to 0} f^\delta_\eta(\tau) = \delta(\tau-\eta) \eqend{,}
\end{equation}
and $g_{\delta'}(\vec{x})$ is a real smooth function of compact support depending on a parameter $\delta'$ such that for its Fourier transform $\tilde{g}_{\delta'}$
we have
\begin{equation}
\lim_{\delta' \to 0} \abs{\tilde{g}_{\delta'}(\vec{p})}^2 = (2\pi)^3 \delta^3(\vec{p}) \eqend{.}
\end{equation}
With this choice of $\hat{u}^{\mu\nu}$ and taking $\hat{w}^{\alpha\mu\beta\nu} = 0$, it then follows that
\begin{equation}
\mathcal{P}_\text{S}^2(\abs{\vec{k}},\eta) = \lim_{\delta \to 0} \lim_{\delta' \to 0} \bra{0} I(f,\chi) I(f^*,\chi^*) \ket{0}
\end{equation}
whenever this limit exists, and a similar (but more elaborate) construction of test functions is needed for the tensor power spectrum. The right-hand side, before taking the limit $\delta, \delta' \to 0$, represents a smeared version of the power spectrum, where $f^\delta_\eta(\tau)$ is a smearing in time around the point $\tau = \eta$ and $g_{\delta'}(\vec{x})$ limits the spatial integration to a finite volume. While for finite smearing scales (\ie, $\delta, \delta' > 0$) this is well defined, the existence of the limit $\delta, \delta' \to 0$ is not guaranteed in general. In particular, while at tree level taking this limit we obtain the well-known finite result~\eqref{power_spectra}, explicit calculations~\cite{froeb_roura_verdaguer_2012} show that once loop corrections are fully taken into account only one of the limits can be taken, and taking both leads to a divergent answer. Various resolutions of this problem have been proposed, re-examining the connection between the observed power spectrum and its definition in terms of inflationary correlation functions~\cite{agullo_etal_prl_2008,durrer_marozzi_rinaldi_prd_2009,miao_woodard_jcap_2012,basterogil_etal_prd_2013,miao_park_prd_2014}. We note, however, that the smearing is unobservable as long as it is small (for the smearing in time) or large enough (for the spatial smearing). In particular, the lowest observable mode, corresponding to the quadrupole in the Cosmic Microwave Background, probes the power spectrum at wavelengths $\abs{\vec{p}}_\text{min} \approx 2/\Delta \eta$, where $\Delta \eta$ is the difference in conformal time between recombination and today~\cite{mukhanov_textbook,dodelson_textbook,baumann_lectures_2009}. By extending the spatial integration [\ie, the support of $g_{\delta'}(\vec{x})$ in the test function~\eqref{u_testfunction_example}] to a large enough region, the power spectrum is unmodified for all $\abs{\vec{p}} \gtrsim \abs{\vec{p}}_\text{min}$.

\section{Discussion}
\label{discussion}

In this paper we showed that the gauge-invariant and compactly supported observables, $I(f,\chi)$, in single-field inflation are equivalent to (compactly smeared) gauge-invariant local operators, with linearised gravity in de~Sitter and Minkowski spaces being special cases. One of these local observables is the linearised Weyl tensor, while the other local observable is given by equation~\eqref{C_expression}, whose geometric meaning is not completely clear. One crucial ingredient in the proof of this fact was that the test tensor $f^{\mu\nu}$ for the compactly supported observable [see equation~\eqref{i_def}] times $a^{n+2}$ can be chosen to be divergence free with respect to the flat metric without changing the observable $I(f,\chi)$, as long as the graviton and inflaton fields are on-shell. We find this result described in appendix~\ref{appendix_onshell_scalar} rather remarkable. For example, in equation~\eqref{remarkable} the coefficient of $t_{\mu} t_\nu f^{\mu\nu}$ is the $\eta$-derivative of the coefficient of $H a f + \phi' \chi$. There seems to be no reason to expect this to happen, but it is a crucial condition one needs because the former has to vanish, whereas the latter has to be a constant. Another seemingly miraculous aspect in equation~\eqref{remarkable} is that the variable $w$ occurs in a total divergence and nowhere else, which is also a crucial condition. We have not found any underlying reason why these simplifications occur. The other crucial ingredient was a lemma, proved in appendix~\ref{appendix_lemma}, that a smooth, compactly supported and divergence-free symmetric tensor $f^{\mu\nu}$ in flat space can be expressed as $f^{\mu\nu} = \partial_\alpha \partial_\beta u^{\alpha\mu\beta\nu}$, where the tensor $u^{\alpha\mu\beta\nu}$ has the symmetry of the Riemann tensor and is also smooth and compactly supported. This lemma is analogous to the Poincar\'e lemma in the de~Rham cohomology with compact support and appears to be known in the mathematics community~\cite{khavkine_jgeomphys_2014}, although we believe that our proof that the support of $u^{\alpha\mu\beta\nu}$ can be chosen close to the support of $f^{\mu\nu}$ (in the sense explained in appendix~\ref{appendix_lemma}) is new.

We then expressed the linearised Weyl tensor and the tensor $C_{\mu\nu}$, whose spacetime smearing is equivalent to $I(f,\chi)$, in terms of the (gauge-invariant) tensor perturbation $H_{kl}$ and the Sasaki--Mukhanov variable $Q$, which carries the scalar perturbation. Since the correlators of $H_{kl}$ and $Q$ are IR divergent and suffer from secular and long-distance growth, it might appear that the strong correlation of these variables at long distances could be detectable in principle. On the other hand, derivatives of these variables are IR finite and do not show secular or long-distance growth. Thus, because we found that these variables always occur with derivatives in $I(f,\chi)$, the correlator for the compactly supported observable $I(f,\chi)$ is also IR finite and lacks secular or long-distance growth. Since the gauge-invariant variables $H_{kl}$ and $Q$ are non-locally related to the graviton and inflaton fields, the local values of $H_{kl}$ or $Q$ cannot be measured in a local measurement. On the other hand, the compactly supported $I(f,\chi)$ models an observable that \emph{can} be measured locally. Thus, our result indicates that observables which can be measured locally are IR finite and not correlated strongly at long distances, unlike the variables $H_{kl}$ and $Q$. In de~Sitter space this weak correlation of local observables may be related to the fact that the IR divergences for gravitons can be gauged away in any arbitrarily large but finite region by a gauge transformation which in this region corresponds to a rescaling of coordinates~\cite{higuchi_marolf_morrison_cqg_2011}. Therefore, it is natural to speculate that the IR divergences for the graviton and inflaton can also be gauged away in the same way in single-field inflation. While our treatment differs from the usual one in cosmology, we have shown that one can nevertheless recover the well-known scalar and tensor power spectra from the correlator of the compactly supported observable $I(f,\chi)$ at all observable scales.

Our analysis does not make any statement about observables which are not of compact support, or about 
large gauge transformations, which are gauge transformations that do not fall off at spatial infinity. In fact, it has been argued~\cite{anninos_ng_strominger_cqg_2011,hinterbichler_hui_khoury_jcap_2012,woodard_jhep_2016,ferreira_etal_arxiv_2016} that those gauge transformations correspond to asymptotic symmetries of the spacetime which change the physical state. Nevertheless, those changes can only be measured by a meta-observer who can compare different asymptotic regions, and are unobservable for any single observer, for whom the compactly supported observable
considered in this paper is relevant.

Finally, let us comment on the fact that the correlator for the field $C_{\mu\nu}$ is singular in the de~Sitter limit $\epsilon \to 0$. This is simply due to the fact that there is some freedom in the choice of $u^{\alpha\mu\beta\nu}$, and a random choice results in $\chi$ that is singular in this limit. However, one is free to choose a pair $(f^{\mu\nu},\chi)$ such that $\chi = \bigo{\epsilon^0}$ and $f = \bigo{\epsilon}$ (recall that in the de~Sitter limit, $f^{\mu\nu}$ can be made traceless on-shell). Also, if one chooses $u^{\alpha\mu\beta\nu}$ to be traceless, then $\chi = 0$. Thus, the singularity of $C_{\mu\nu}$ in the de~Sitter limit is not an intrinsic property of the compactly supported observable $I(f,\chi)$.

\acknowledgments

We thank Chris Fewster for useful discussions in an early stage of this work, and Igor Khavkine for discussions on ref.~\cite{canepa_dappiaggi_khavkine_2017}. This work is part of a project that has received funding from the European Union’s Horizon 2020 research and innovation programme under the Marie Sk{\l}odowska-Curie grant agreement No. 702750 ``QLO-QG''.

\appendix

\section{Conformal transformation}
\label{appendix_conformal}

Under the conformal transformation
\begin{equation}
\tilde{g}_{\mu\nu} = a^2 g_{\mu\nu}
\end{equation}
the Christoffel symbols transform as
\begin{equation}
\tilde{\Gamma}^\alpha_{\mu\nu} = \Gamma^\alpha_{\mu\nu} + a^{-1} \left( \delta^\alpha_\mu \delta^\sigma_\nu + \delta^\alpha_\nu \delta^\sigma_\mu -
g_{\mu\nu} g^{\alpha\sigma} \right) \partial_\sigma a \eqend{,}
\end{equation}
and the curvature tensors become
\begin{subequations}
\begin{align}
\label{app_riemann}
\tilde{R}^\alpha{}_{\beta\gamma\delta} &= R^\alpha{}_{\beta\gamma\delta} - 2 a^{-2} \delta^\alpha_{[\gamma} g_{\delta]\beta} ( \nabla a )^2 + 4 a^{-2}
g^{\alpha\tau} \delta^\sigma_{[\gamma} g_{\delta][\tau} \left[ a \nabla_{\beta]} \nabla_\sigma a - 2 ( \nabla_{\beta]} a ) ( \nabla_\sigma a ) \right] \\
\begin{split}
\tilde{R}_{\mu\nu} &= R_{\mu\nu} - (n-2) a^{-1} \nabla_\mu \nabla_\nu a + 2 (n-2) a^{-2} ( \nabla_\mu a ) ( \nabla_\nu a ) \\
&\quad- g_{\mu\nu} \left[ (n-3) a^{-2} ( \nabla a )^2 + a^{-1} \nabla^2 a \right]
\end{split} \\
a^2 \tilde{R} &= R - (n-1) \left[ 2 a^{-1} \nabla^2 a + (n-4) a^{-2} ( \nabla a )^2 \right] \eqend{,}
\end{align}
\end{subequations}
where $\nabla_\mu$ is the covariant derivative associated with $g_{\mu\nu}$, $\nabla^2 = \nabla^\mu \nabla_\mu$, and where we used the notation
\begin{equation}
( \nabla a )^2 \equiv ( \nabla^\mu a ) ( \nabla_\mu a ) \eqend{.}
\end{equation}

\section{A Poincar{\'e}-type lemma}
\label{appendix_lemma}

\textit{Given a smooth, symmetric and divergence-free tensor $f^{\mu\nu}$ on $\mathbb{R}^n$, $n\geq 2$, with compact support, \ie, $f^{\mu\nu} = f^{\nu\mu}$, $\partial_\mu f^{\mu\nu} = 0$, and $\operatorname{\supp} f^{\mu\nu} \equiv \Omega_f$, with $\Omega_f = \overline{\Omega}_f$ compact, there exists a smooth tensor $u^{\alpha\mu\beta\nu}$ with the symmetries of the Riemann tensor, \ie, $u^{\alpha\mu\beta\nu} = u^{\beta\nu\alpha\mu} = u^{[\alpha\mu][\beta\nu]}$ and $u^{\alpha[\mu\beta\nu]} = 0$, such that
\begin{equation}
\label{f_in_terms_of_u}
f^{\mu\nu} = \partial_\alpha \partial_\beta u^{\alpha\mu\beta\nu} \eqend{.}
\end{equation}
The support of $u^{\alpha\mu\beta\nu}$, denoted by
\begin{equation}
\label{support_of_u}
\operatorname{\supp} u^{\alpha\mu\beta\nu} \equiv \Omega_u = \overline{\Omega}_u \eqend{,}
\end{equation}
can be chosen to be any compact star-shaped region in which $\Omega_f$ is strictly contained,
\begin{equation}
\label{omega_f_u_contained}
\Omega_f \subset \Omega_u \setminus \partial\Omega_u \eqend{.}
\end{equation}
}

\proof The construction is different for $n = 2$ and $n > 2$. Let us thus first take $n = 2$. Define
\begin{equation}
u^{1212}(x^1,x^2) \equiv - \int_{-\infty}^{x^1} \left[ \int_{-\infty}^{x^2} f^{12}(y^1,y^2) \total y^2 \right] \total y^1 \eqend{,}
\end{equation}
and define the other components of $u^{\mu\alpha\nu\beta}$ imposing the symmetries of the Riemann tensor. Then one can readily verify that equation~\eqref{f_in_terms_of_u} is satisfied.  Next, since the second-order partial derivatives of $u^{1212}$ are given by $f^{\mu\nu}$, on any straight line $x(\lambda)$ that does not intersect $\Omega_f$ we have
\begin{equation}
\frac{\total^2}{\total \lambda^2} u^{1212}(x(\lambda)) = 0 \eqend{.}
\end{equation}
This implies that $u^{1212}(x(\lambda)) = 0$ on any piecewise linear curve starting from infinity (such that $x^k(\lambda) \to -\infty$ as $\lambda \to -\infty$ for $k = 1,2$) which does not intersect $\Omega_f$, since as $x^k \to -\infty$ the zeroth and first $\lambda$ derivatives of $u^{1212}(x(\lambda))$ vanish by construction. Hence the support $\Omega_u$ of $u^{\mu\alpha\nu\beta}$ is the smallest simply-connected closed set containing $\Omega_f$, and equation~\eqref{support_of_u} is also satisfied.

Next we take $n \geq 3$. Without loss of generality we may assume that $\Omega_f$ is already star-shaped, and we choose coordinates such that $\Omega_f$ is star-like with the base point at $x_0 = 0$. Since the result is independent of the signature of spacetime, we in fact work in $n$-dimensional Euclidean space. We first construct a solution $U^{\mu\alpha\nu\beta}$ of the equation $\partial_\alpha \partial_\beta U^{\mu\alpha\nu\beta} = f^{\mu\nu}$ which is not necessarily compactly supported. Define the tensor $F^{\mu\nu}$ to be the solution (with vanishing boundary conditions at infinity) to
\begin{equation}
\laplace F^{\mu\nu} = f^{\mu\nu} \eqend{,}
\end{equation}
which can explicitly be given as
\begin{equation}
\label{def_of_F}
F^{\mu\nu}(x) \equiv - \frac{\Gamma\left( \frac{n}{2} - 1 \right)}{4 \pi^\frac{n}{2}} \int \frac{f^{\mu\nu}(y)}{\abs{x-y}^{n-2}} \total^n y \eqend{.}
\end{equation}
It can readily be verified that $\partial_\mu F^{\mu\nu} = \partial_\nu F^{\mu\nu} = 0$. Thus, the tensor
\begin{equation}
U^{\alpha\mu\beta\nu} \equiv \delta^{\alpha\beta} F^{\mu\nu} - \delta^{\alpha\nu} F^{\mu\beta} - \delta^{\mu\beta} F^{\alpha\nu} + \delta^{\mu\nu}
F^{\alpha\beta}
\end{equation}
has the symmetries of the Riemann tensor (including $U^{\alpha[\mu\beta\nu]} = 0$) and satisfies
\begin{equation}
\label{non-compact_U}
\partial_\alpha \partial_\beta U^{\alpha\mu\beta\nu} = f^{\mu\nu} \eqend{,}
\end{equation}
but is not necessarily of compact support.

For later purposes it is useful to establish the behaviour of $F^{\mu\nu}$, hence of $U^{\alpha\mu\beta\nu}$, as $r \to \infty$. We first note that $\partial_\mu f^{\mu\nu} = 0$ implies $f^{\mu\nu} = \partial_\alpha v^{\alpha\mu\nu}$ for some compactly supported tensor $v^{\alpha\mu\nu} = v^{[\alpha\mu]\nu}$ by the Poincar{\'e} lemma with compact support (see appendix~\ref{appendix_lemma2} for a proof). Thus we have
\begin{equation}
\label{integral_zero}
\int f^{\mu\nu}(y) \total^n y = \int \partial_\alpha v^{\alpha\mu\nu}(y) \total^n y = 0 \eqend{.}
\end{equation}
From equation~\eqref{integral_zero} and the definition~\eqref{def_of_F} we immediately obtain
\begin{equation}
F^{\mu\nu}(x) = - \frac{\Gamma\left( \frac{n}{2} - 1 \right)}{4 \pi^\frac{n}{2}} \int \left[ \frac{1}{\abs{x-y}^{n-2}} - \frac{1}{r^{n-2}}\right] f^{\mu\nu}(y) \total^n y \eqend{,}
\end{equation}
where $r \equiv \abs{x}$. Since $\abs{x-y}^{-(n-2)} - r^{-(n-2)}$ decays like $r^{-(n-1)}$ as $r \to \infty$, the tensors $F^{\mu\nu}$ and $U^{\alpha\mu\beta\nu}$ decay like $r^{-(n-1)}$ at infinity, and a similar argument shows that the tensors $\partial_\gamma F^{\mu\nu}$ (and thus $\partial_\gamma U^{\alpha\mu\beta\nu}$) decay like $r^{-n}$ at infinity.

Now our task is to subtract a term from $U^{\alpha\mu\beta\nu}$ in such a way that the resulting tensor $u^{\alpha\mu\beta\nu}$ has compact support while still satisfying equation~\eqref{f_in_terms_of_u}. We choose a star-shaped region $\Omega_u$ satisfying the condition~\eqref{omega_f_u_contained}, and an auxiliary function $\chi$ which is smooth and satisfies
\begin{equation}
\label{appendix_lemma_chi}
\chi(x) = \begin{cases} 0 & x \in \Omega_f \\ 1 & x \in \mathbb{R}^n \setminus \Omega_u \eqend{.} \end{cases}
\end{equation}
We also define the integral operators $\mathcal{G}_k$ on any function $f$ by
\begin{equation}
\label{def_G_k}
\left( \mathcal{G}_k f \right)(x) \equiv \int_1^\infty t^{k-1} f(t x) \total t \eqend{.}
\end{equation}
By letting $\rho \equiv t r$, we can write these operators as
\begin{equation}
(\mathcal{G}_k f)(x) = r^{-k} \int_r^\infty \rho^{k-1} f(\rho x/\abs{x}) \total \rho \eqend{.}
\end{equation}
In this form it is clear that the operators $\mathcal{G}_k$ are well defined for $r > 0$ on functions which decay faster than $r^{-k}$ at infinity, and map functions with star-shaped compact support to functions with the same support. We note that, if $f$ decays as fast as $r^{-l}$ at infinity, \ie, if $\abs{f(r)} \leq c r^{-l}$ for $r \geq R$ for some $R$, where $c$ is a positive constant, then for all $r \geq R$ and all $k < l$ we have
\begin{equation}
\abs{\mathcal{G}_k f} \leq r^{-k} \int_r^\infty \rho^{k-1} \abs{f(\rho)} \total \rho \leq \frac{c}{l-k} r^{-l} \eqend{.}
\end{equation}
Thus, $\mathcal{G}_k f$ also decays as fast as $r^{-l}$ at infinity. Using the definition~\eqref{def_G_k} and the equality $x^\alpha \partial_\alpha f(t x) = t \partial_t f(t x)$, one readily finds that
\begin{equation}
\label{appendix_lemma_g_prop1}
\left( x^\alpha \partial_\alpha + k \right) \mathcal{G}_k f = - f \eqend{,}
\end{equation}
and that
\begin{equation}
\label{appendix_lemma_g_prop2}
\partial_\alpha \mathcal{G}_k f = \mathcal{G}_{k+1} \partial_\alpha f
\end{equation}
if $\partial_\alpha f$ decays faster than $r^{-(k+1)}$ at infinity.

Define now the tensors
\begin{equations}
\bar{U}^{\alpha\mu\beta\nu} &\equiv \mathcal{G}_{n-2} U^{\alpha\mu\beta\nu} \eqend{,} \\
\bar{\bar{U}}^{\alpha\mu\beta\nu} &\equiv \mathcal{G}_{n-3} \bar{U}^{\alpha\mu\beta\nu} \eqend{.}
\end{equations}
These tensors are well defined for $r > 0$ because the tensor $U^{\alpha\beta\mu\nu}$ decays like $r^{-(n-1)}$ as $r \to \infty$, as we have shown
before, \ie, they decay faster than $r^{-(n-2)}$. They also inherit the symmetries of $U^{\alpha\mu\beta\nu}$. Then we define
\begin{equation}
\label{appendix_lemma_w}
W^{\alpha\mu\beta\nu\rho} \equiv x^\rho \bar{U}^{\alpha\mu\beta\nu} + V^{\alpha\mu\beta\nu\rho} - V^{\mu\alpha\beta\nu\rho} + V^{\beta\nu\alpha\mu\rho} - V^{\nu\beta\alpha\mu\rho}
\end{equation}
with
\begin{equation}
\label{appendix_lemma_v}
V^{\alpha\mu\beta\nu\rho} \equiv \frac{1}{2} x^\alpha \left( \bar{U}^{\rho\mu\nu\beta} + x^\nu \partial_\sigma \bar{\bar{U}}^{\rho(\beta\mu)\sigma} - x^\beta \partial_\sigma \bar{\bar{U}}^{\rho(\mu\nu)\sigma} + x^\rho \partial_\sigma \bar{\bar{U}}^{\sigma\mu\nu\beta} \right) \eqend{.}
\end{equation}
Note that the tensor $V^{\alpha\mu\beta\nu\rho}$ is antisymmetric under the exchange $\beta \leftrightarrow \nu$ by construction. It thus follows that the tensor $W^{\alpha\mu\beta\nu\rho}$ is antisymmetric under each of the exchanges $\alpha \leftrightarrow \mu$ and $\beta \leftrightarrow \nu$ and symmetric under the pairwise exchange $(\alpha\mu) \leftrightarrow (\beta\nu)$. We also find by an explicit calculation that $V^{\alpha[\mu\beta\nu]\rho} = 0$, and then also $W^{\alpha[\mu\beta\nu]\rho} = 0$. Therefore, the tensor $W^{\alpha\mu\beta\nu\rho}$ has the symmetry of the Riemann tensor in the first four indices. We can also organise this tensor as follows:
\begin{splitequation}
\label{W_sym_manifest}
W^{\alpha\mu\beta\nu\rho} &= x^{[\rho} \bar{U}^{\alpha]\mu\beta\nu} + x^{[\rho} \bar{U}^{\beta]\nu\alpha\mu} - x^\mu x^\nu \partial_\sigma \bar{\bar{U}}^{\rho(\alpha\beta)\sigma} + \frac{1}{2}\left( x^\alpha x^\nu \partial_\sigma \bar{\bar{U}}^{\rho\beta\mu\sigma} + x^\beta x^\mu \partial_\sigma \bar{\bar{U}}^{\rho\alpha\nu\sigma}\right) \\
&\quad+ x^\nu x^{[\alpha} \partial_\sigma \bar{\bar{U}}^{\rho]\mu\beta\sigma} + x^\mu x^{[\beta}\partial_\sigma \bar{\bar{U}}^{\rho]\nu\alpha\sigma} - x^\alpha x^{[\beta} \partial_\sigma \bar{\bar{U}}^{\rho]\nu\mu\sigma} - x^\beta x^{[\alpha} \partial_\sigma \bar{\bar{U}}^{\rho]\mu\nu\sigma} \eqend{.}
\end{splitequation}
In this form it is clear that the tensor $W^{\alpha\mu\beta\nu\rho}$ is a sum of terms which are antisymmetric under one of the exchanges $\alpha \leftrightarrow \rho$ and $\beta \leftrightarrow \rho$.

We then obtain after a long but straightforward calculation using the properties~\eqref{appendix_lemma_g_prop1} and~\eqref{appendix_lemma_g_prop2} of the integral operators that
\begin{equation}
\partial_\rho V^{\alpha\mu\beta\nu\rho} = - \frac{1}{2} \bar{U}^{\alpha\mu\beta\nu} + \frac{3}{4} x^\alpha \partial_\rho \bar{\bar{U}}^{\beta\nu\mu\rho}
+ \frac{1}{2} x^\nu \partial_\rho \bar{\bar{U}}^{\alpha(\beta\mu)\rho} - \frac{1}{2} x^\beta \partial_\rho \bar{\bar{U}}^{\alpha(\mu\nu)\rho} - x^\alpha
x^{[\beta} \mathcal{G}_{n-1} \mathcal{G}_n f^{\nu]\mu}\eqend{.}
\end{equation}
From this it follows that
\begin{equation}
\partial_\rho W^{\alpha\mu\beta\nu\rho} = - U^{\alpha\mu\beta\nu} + 4 x^{[\alpha} \left( \mathcal{G}_{n-1} \mathcal{G}_n f^{\mu][\beta} \right) x^{\nu]}
\eqend{,}
\end{equation}
and finally we define
\begin{equation}
u^{\alpha\mu\beta\nu} \equiv U^{\alpha\mu\beta\nu} + \partial_\rho \left( \chi \, W^{\alpha\mu\beta\nu\rho} \right)
\end{equation}
with the auxiliary function $\chi$ specified by equation~\eqref{appendix_lemma_chi}. Note that, although the tensor $W^{\alpha\mu\beta\nu\rho}$ is undefined at $r = 0$, the tensor $u^{\alpha\mu\beta\nu}$ is well defined there as well because $\chi(x) = 0$ if $x \in \Omega_f$, which contains the origin. The tensor $u^{\alpha\mu\beta\nu}$ clearly has the symmetries of the Riemann tensor. Equation~\eqref{f_in_terms_of_u} follows from equation~\eqref{non-compact_U} and the antisymmetry properties of each term in $W^{\alpha\mu\beta\nu\rho}$ manifest in equation~\eqref{W_sym_manifest}, which imply
\begin{equation}
\partial_\alpha \partial_\beta \partial_\rho \left( \chi \, W^{\alpha\mu\beta\nu\rho} \right) = 0 \eqend{.}
\end{equation}
Finally, the tensor $u^{\alpha\mu\beta\nu}$ has support in $\Omega_u$ because for $x \in \mathbb{R}^n \setminus \Omega_u$, where $f^{\mu\nu} = 0$ and $\chi = 1$, we have $\partial_\rho W^{\alpha\mu\beta\nu\rho} = - U^{\alpha\mu\beta\nu}$ and hence $u^{\alpha\mu\beta\nu} = 0$.
\endproof

\section{The Poincar{\'e} lemma for vectors with compact support}
\label{appendix_lemma2}

\textit{Given a smooth and divergence-free vector $f^\mu$ on $\mathbb{R}^n$, $n \geq 2$, with compact support, there exists a smooth antisymmetric tensor $u^{\mu\nu}$ such that $f^\nu = \partial_\mu u^{\mu\nu}$. The support of $u^{\mu\nu}$ can be chosen to be any compact star-shaped region in which the support of $f^\mu$ is strictly contained.}

\proof The construction proceeds in close analogy to the lemma proved in appendix~\ref{appendix_lemma}, and we again distinguish the cases $n = 2$ and $n > 2$. For $n = 2$, we find
\begin{equation}
u^{12}(x^1,x^2) = \int_{-\infty}^{x^1} f^2(y,x^2) \total y \eqend{.}
\end{equation}
One can show that the support of $u^{12}$ is the smallest simply-connected closed set containing the support of $f^\mu$ by noting that on any piecewise smooth curve $x^\mu(\lambda)$ outside the support of $f^\mu$ one has $\total u^{12}(x(\lambda))/\total \lambda = 0$.

Next we let $n \geq 3$. We first construct a solution to the equation $\laplace A^\mu = f^\mu$ as
\begin{equation}
A^\mu \equiv - \frac{\Gamma\left( \frac{n}{2}-1 \right)}{4\pi^{\frac{n}{2}}} \int \frac{f^\mu(y)}{\abs{x-y}^{n-2}} \total^n y \eqend{.}
\end{equation}
Since $\partial_\mu A^\mu = 0$, the tensor $U^{\mu\nu} \equiv \partial^\mu A^\nu - \partial^\nu A^\mu$ satisfies $\partial_\mu U^{\mu\nu} = f^\nu$ but is not necessarily compactly supported. It can be shown as in appendix~\ref{appendix_lemma} that $U^{\mu\nu}$ decays like $r^{-n}$ at infinity. Then the tensor $u^{\mu\nu} \equiv U^{\mu\nu} + 3 \partial_\rho \left( \chi \, x^{[\rho} \mathcal{G}_{n-2} U^{\mu\nu]} \right)$ can be shown to satisfy all the required properties in the same way as in appendix~\ref{appendix_lemma}.
\endproof

\noindent\textbf{Remark}. If we replace $f^\mu$ by a compactly supported symmetric tensor $f^{\mu\nu}$ satisfying $\partial_\mu \partial_\nu f^{\mu\nu} = 0$ in this lemma, then we can find a compactly supported tensor $u^{\rho\mu\nu}$ antisymmetric in the first two indices such that $f^{\mu\nu} = \partial_\rho u^{\rho\mu\nu} + \partial_\rho u^{\rho\nu\mu}$. This can be proved as follows: Since $g^\mu \equiv \partial_\nu f^{\mu\nu}$ is divergence free, $\partial_\mu g^\mu = 0$, according to the above we can find a compactly supported antisymmetric tensor $q^{\mu\nu}$ such that $g^\mu = \partial_\nu q^{\mu\nu}$. Then $a^{\mu\nu} \equiv f^{\mu\nu} - q^{\mu\nu}$ is divergence free in the second index, $\partial_\nu a^{\mu\nu} = 0$. Hence according to the above we have $a^{\mu\nu} = 2 \partial_\rho u^{\mu\rho\nu}$, where $u^{\mu\rho\nu}$ is compactly supported and antisymmetric in the last two indices. Symmetrising this equation in $\mu$ and $\nu$, we obtain the required result.

\section{On-shell elimination of a scalar-type linear combination of test functions}
\label{appendix_onshell_scalar}

In this appendix we prove equations~\eqref{check_equations}, which are equivalent to $H a \delta f^{\mu\nu} + \phi' \delta \chi = - H a f^{\mu\nu} - \phi' \chi$, which, in turn, is equivalent to
\begin{equation}
\label{miracle}
S(v,w) \equiv g^{\mu\nu} E_{\mu\nu}^{(1)}(v,w) - \kappa^2 \frac{\phi'}{H a} F^{(1)}(v,w) = - f - \frac{\phi'}{Ha} \chi \eqend{,}
\end{equation}
where $v^{\mu\nu}$ and $w$ are given by equations~\eqref{w_v_expressions}. Instead of simply substituting these equations and verifying equation~\eqref{miracle}, we start with the following general ansatz for $v^{\mu\nu}$:
\begin{equation}
\label{v_form}
v^{\mu\nu}(x) = A(\eta) \left( f^{\mu\nu}(x) - \frac{1}{n-1} g^{\mu\nu} f(x) \right) + g^{\mu\nu} \alpha(x) \eqend{.}
\end{equation}
We also find it instructive to define
\begin{equation}
\label{S_different}
S(v,w,z) \equiv g^{\mu\nu} E_{\mu\nu}^{(1)}(v,w) - \kappa^2 z(\eta) F^{(1)}(v,w) \eqend{,}
\end{equation}
and let $z = \phi'/(H a)$ so that $S(v,w,z) = S(v,w)$ in the end.

By substituting equation~\eqref{v_form} into \eqref{S_different} we immediately find that, in order not to have any second derivative of $w$ in $S(v,w,z)$, we need to set
\begin{equation}
\label{def_alpha}
\alpha = \kappa^2 \frac{z}{(n-1) (n-2)} w \eqend{.}
\end{equation}
By substituting equation~\eqref{v_form} with this form of $\alpha$ into equation~\eqref{S_different} we find
\begin{splitequation}
\label{messy_S}
S &= \nabla_\rho \left[ \kappa^2 \left[ 2 \nabla^\rho z + \left( n-2 - \frac{\kappa^2 z^2}{2 (n-1)} \right) \nabla^\rho \phi \right] w - \left( (n-2) A \chi + \kappa^2 \frac{z}{2 (n-1)} A f \right) \nabla^\rho \phi \right] \\
&\quad+ \kappa^2 \left[ - \nabla^\rho \nabla_\rho z + \left( 1 + \kappa^2 \frac{n z^2}{4 (n-1) (n-2)} \right) V'(\phi) + \kappa^2 \frac{n z}{2 (n-1) (n-2)} V(\phi) + \frac{z V''(\phi)}{2} \right. \\
&\qquad\qquad \left. + \kappa^2 \frac{z}{2 (n-1)} \nabla_\rho z \nabla^\rho \phi \right] w + \left[ - (n-2) \nabla_\mu \nabla_\nu A + \kappa^2 z \nabla_\mu \phi \nabla_\nu A + \kappa^2 z A \nabla_\mu \nabla_\nu \phi \right] f^{\mu\nu} \\
&\quad+ \left[ - (n-2) \nabla_\mu A \nabla^\mu \phi + \kappa^2 z A \nabla^\mu \phi \nabla_\mu \phi  \right] \chi \\
&\quad+ \left[ - \frac{\kappa^2}{2 (n-1)} V(\phi) + \frac{\kappa^2}{2 (n-1)} \nabla^\mu z \nabla_\mu \phi  - \kappa^2 \frac{z}{4 (n-1)} V'(\phi) \right] A f \eqend{.}
\end{splitequation}
Note that this equation is valid in any spacetime. In FLRW spacetime, upon substitution of $z = \phi'/(H a)$ with
\begin{equations}
z' &= - (n-1-\epsilon) \phi' - \frac{a}{2 H} V'(\phi) \eqend{,} \\
z'' &= - (n-1-\epsilon) \phi'' + \epsilon' \phi' - \frac{a}{2 H} \phi' V''(\phi) - \frac{1+\epsilon}{2} a^2 V'(\phi) \eqend{,}
\end{equations}
equation~\eqref{messy_S} simplifies remarkably to
\begin{splitequation}
\label{remarkable}
S &= - \frac{1}{H a} \nabla_\rho \left[ \kappa^2 H t^\rho \left[ 2 z' + \left( n-2 - \kappa^2 \frac{z^2}{2 (n-1)} \right) \phi' \right] w - H t^\rho \phi' \left( (n-2) A \chi + \kappa^2 \frac{z}{2 (n-1)} A f \right) \right] \\
&\quad+ \frac{n-2}{H a} \left[ - \left( \frac{H}{a} A \right)'' t_\mu t_\nu f^{\mu\nu} + \left( \frac{H}{a} A \right)' \left( H a f + \phi' \chi \right) \right] \eqend{.}
\end{splitequation}
Hence we achieve
\begin{equation}
S = - f - \frac{\phi'}{H a} \chi
\end{equation}
by requiring that $(H A/a)' = - 1/(n-2)$ and that the expression inside the brackets in the first line vanishes. That is,
\begin{equation}
w = A \frac{2 (n-1) (n-2) \chi + \kappa^2 z f}{\kappa^2 \left[ 4 (n-1) z' / \phi' + 2 (n-2) (n-1) - \kappa^2 z^2 \right]} \eqend{,}
\end{equation}
which can be written in the form~\eqref{w_expression}, with $A(\eta)$ given by equation~\eqref{def_A_eta}.

\noindent\textbf{Remark}. If we let $A(\eta) = 0$ and $z = \sqrt{2 (n-1) (n-2)} / \kappa$ in equation~\eqref{messy_S}, then we have
\begin{equation}
S = \frac{w}{\kappa \sqrt{2 (n-1) (n-2)}} \left[ n \kappa^2 V(\phi) + \sqrt{2 (n-1) (n-2)} \left( 1 + \frac{n}{2} \right) \kappa V'(\phi) + (n-1) (n-2) V''(\phi) \right] \eqend{.}
\end{equation}
Then one can choose $w$ such that $S = - f - \kappa^2 z \chi$. This means that it is possible to set $\check{f} + \kappa^2 z \check{\chi} = 0$ if $z = \sqrt{2 (n-1) (n-2)}/\kappa$ for any generic spacetime generated by a scalar field with potential $V(\phi)$, although we have not found any application of this fact.

\section{Expressing the tensor $C_{\mu\nu}$ as a geometric quantity}
\label{appendix_geometric}

In this appendix we show that all components of the tensor $C_{\mu\nu}$ can be expressed as a linear perturbation of geometric quantities, except for $C_{00}$. It is convenient to introduce the following expressions, where a bar over a tensor indicates its projection on each free index onto spatial coordinates, \eg,
\begin{equations}
\bar{\partial}_\mu &\equiv \left( \delta_\mu^\nu - \delta_\mu^0 \delta_0^\nu \right) \partial_\nu = \partial_\mu - \delta_\mu^0 \partial_\eta \eqend{,} \\
\bar{\eta}_{\mu\nu} &\equiv \left( \delta_\mu^\alpha - \delta_\mu^0 \delta^\alpha_0 \right) \left( \delta_\nu^\beta - \delta_\nu^0 \delta^\beta_0 \right) \eta_{\alpha\beta} = \eta_{\mu\nu} + \delta_\mu^0 \delta_\nu^0 \eqend{,} \\
\bar{h}_{\mu\nu} &\equiv h_{\mu\nu} - 2 \delta^0_{(\mu} h_{\nu)0} + \delta^0_\mu \delta^0_\nu h_{00} \eqend{.}
\end{equations}
It is also useful to define an invariant four-velocity by the normalised gradient of the scalar field
\begin{equation}
\tilde{u}_\mu \equiv \frac{\nabla_\mu \tilde{\phi}}{\sqrt{- \tilde{g}^{\rho\sigma} \nabla_\rho \tilde{\phi} \nabla_\sigma \tilde{\phi}}} \eqend{,}
\end{equation}
which has the following background value:
\begin{equation}
u_\mu = - t_\mu = a \delta^0_\mu \eqend{.}
\end{equation}
Its part linear in perturbations is given by
\begin{equation}
u^{(1)}_\mu = - \frac{a}{2} \delta^0_\mu h_{00} + \frac{a}{\phi'} \bar{\partial}_\mu \phi^{(1)} \eqend{.}
\end{equation}
From this, we define the acceleration as
\begin{equation}
\tilde{a}_\mu \equiv \tilde{u}^\nu \tilde{\nabla}_\nu \tilde{u}_\mu \eqend{,}
\end{equation}
which has vanishing background value and linear perturbation
\begin{equation}
a^{(1)}_\mu = - \frac{1}{2} \bar{\partial}_\mu h_{00} - \frac{1}{H a} \partial_\eta \bar{\partial}_\mu \left( \frac{H a}{\phi'} \phi^{(1)} \right) - \epsilon \bar{\partial}_\mu \left( \frac{H a}{\phi'} \phi^{(1)} \right) \eqend{.}
\end{equation}
We also define the projector
\begin{equation}
\tilde{\Pi}_\mu^\nu \equiv \delta_\mu^\nu + \tilde{u}_\mu \tilde{u}^\nu \eqend{,}
\end{equation}
whose background value reads
\begin{equation}
\Pi_\mu^\nu = \delta_\mu^\nu - \delta^0_\mu \delta_0^\nu = \bar{\delta}_\mu^\nu \eqend{.}
\end{equation}
Its part linear in perturbation is
\begin{equation}
\Pi^{(1)}_\mu{}^\nu = - \delta^0_\mu h^{0\nu} + a^{-1} \delta^0_\mu \eta^{\nu\rho} u^{(1)}_\rho - a^{-1} \delta_0^\nu u^{(1)}_\mu \eqend{.}
\end{equation}

Next we identify useful geometric quantities. The extrinsic curvature tensor of the hypersurfaces on which the scalar field $\tilde{\phi}$ is constant (the constant-field hypersurface) is
\begin{equation}
\tilde{K}_{\mu\nu} \equiv \tilde{\Pi}_\mu^\rho \tilde{\nabla}_\rho \tilde{u}_\nu = \tilde{\nabla}_\mu \tilde{u}_\nu + \tilde{u}_\mu \tilde{u}^\rho \tilde{\nabla}_\rho \tilde{u}_\nu \eqend{.}
\end{equation}
Its background value and linear part are
\begin{equation}
K_{\mu\nu} = - H a^2 \left( \eta_{\mu\nu} + \delta^0_\mu \delta^0_\nu \right)
\end{equation}
and
\begin{splitequation}
K^{(1)}_{\mu\nu} &= \frac{1}{H} \bar{\partial}_\mu \bar{\partial}_\nu \left( \frac{H a}{\phi'} \phi^{(1)} \right) - 2 a \delta^0_{(\mu} \bar{\partial}_{\nu)} \left( \frac{H a}{\phi'} \phi^{(1)} \right) \\
&\quad+ a \partial_{(\mu} \left( h_{\nu)0} - \delta^0_{\nu)} h_{00} \right) - \frac{a}{2} \partial_\eta \left( h_{\mu\nu} - \delta^0_\mu \delta^0_\nu h_{00} \right) - \frac{H a^2}{2} \left( 2 h_{\mu\nu} + \eta_{\mu\nu} h_{00} - \delta^0_\mu \delta^0_\nu h_{00} \right) \eqend{,}
\end{splitequation}
respectively.
Since $K_{\mu\nu} \neq 0$, $K^{(1)}_{\mu\nu}$ is not gauge invariant. However, the combination
\begin{equation}
\tilde{L}_{\mu\nu} \equiv - 2 \left[ \tilde{K}_{\mu\nu} - \frac{1}{(n-1)} \left( \tilde{g}_{\mu\nu} + \tilde{u}_\mu \tilde{u}_\nu \right) \tilde{\nabla}_\alpha \tilde{u}^\alpha \right] \tilde{\nabla}_\beta \tilde{u}^\beta
\end{equation}
has vanishing background value. Hence its first-order part is gauge invariant and given by
\begin{splitequation}
L^{(1)}_{\mu\nu} &= 2 \left[ (n-1) \bar{\partial}_\mu \bar{\partial}_\nu - \bar{\eta}_{\mu\nu} \laplace \right] \left( \frac{H a}{\phi'} \phi^{(1)} \right) \\
&\quad+ 2 (n-1) H a \bar{\delta}_{(\mu}^\alpha \bar{\partial}_{\nu)} h_{\alpha0} - (n-1) H a \partial_\eta \bar{h}_{\mu\nu} - \bar{\eta}_{\mu\nu} \left[ H a \partial_\eta \left( h_{00} - h \right) + 2 H a \partial^\rho h_{0\rho} \right] \eqend{.}
\end{splitequation}

We also construct the curvature tensor $\tilde{\mathcal{R}}_{\alpha\beta\gamma\delta}$ of the constant-field hypersurfaces, which is related to the spacetime Riemann tensor $\tilde{R}_{\alpha\mu\beta\nu}$ through the Gau{\ss}-Codazzi equation as
\begin{equation}
\tilde{\mathcal{R}}_{\alpha\beta\gamma\delta} = \tilde{\Pi}^\mu_\alpha \tilde{\Pi}^\nu_\beta \tilde{\Pi}^\rho_\gamma \tilde{\Pi}^\sigma_\delta \tilde{R}_{\mu\nu\rho\sigma} - \tilde{K}_{\alpha\gamma} \tilde{K}_{\beta\delta} + \tilde{K}_{\alpha\delta} \tilde{K}_{\beta\gamma} \eqend{.}
\end{equation}
Its background value vanishes, and the part linear in perturbation, which is thus gauge invariant, reads
\begin{equation}
\mathcal{R}^{(1)}_{\alpha\beta\gamma\delta} = - a^2 \bar{\partial}_\gamma \bar{\partial}_{[\alpha} \bar{h}_{\beta]\delta} + a^2 \bar{\partial}_\delta \bar{\partial}_{[\alpha} \bar{h}_{\beta]\gamma} - 4 a^2 \bar{\partial}_{[\alpha} \bar{\eta}_{\beta][\gamma} \bar{\partial}_{\delta]} \left( \frac{H a}{\phi'} \phi^{(1)} \right) \eqend{.}
\end{equation}
We then obtain the first-order part of the Ricci tensor and Ricci scalar of the constant-field hypersurfaces as
\begin{equation}
\mathcal{R}^{(1)}_{\mu\nu} = \bar{\partial}^\beta \bar{\partial}_{(\mu} \bar{h}_{\nu)\beta} - \frac{1}{2} \bar{\partial}_\mu \bar{\partial}_\nu \left( h + h_{00} \right) - \frac{1}{2} \laplace \bar{h}_{\mu\nu} + \left[ (n-3) \bar{\partial}_\mu \bar{\partial}_\nu + \bar{\eta}_{\mu\nu} \laplace \right] \left( \frac{H a}{\phi'} \phi^{(1)} \right)
\end{equation}
and
\begin{equation}
\label{appendix_geometric_ricci3}
a^2 \mathcal{R}^{(1)} = \bar{\partial}^\alpha \bar{\partial}^\beta \bar{h}_{\alpha\beta} - \laplace \left( h + h_{00} \right) + 2 (n-2) \laplace \left( \frac{H a}{\phi'} \phi^{(1)} \right) \eqend{,}
\end{equation}
respectively. We also need to expand the linearised equations of motion~\eqref{eom_linear_e} in the perturbation using the flat-space variable $h_{\mu\nu}$. We find
\begin{splitequation}
E^{(1)}_{\mu\nu} &= 2 R^\text{flat}_{\mu\nu} - \eta_{\mu\nu} R^\text{flat} - 2 (n-2) H a \partial_{(\mu} h_{\nu)0} + (n-2) H a h_{\mu\nu}' - (n-2) H a \eta_{\mu\nu} h' \\
&\quad+ 2 (n-2) H a \eta_{\mu\nu} \partial^\rho h_{0\rho} - (n-2) H^2 a^2 (n-1-\epsilon) \eta_{\mu\nu} h_{00} \\
&\quad- 2 (n-2) \epsilon H a \left( \eta_{\mu\nu} \partial_\eta + 2 \delta^0_{(\mu} \partial_{\nu)} \right) \left( \frac{H a}{\phi'} \phi^{(1)} \right) \\
&\quad+ \left( \eta_{\mu\nu} + 2 \delta^0_\mu \delta^0_\nu \right) (n-1-\epsilon) H a \phi' \phi^{(1)} + \left( \eta_{\mu\nu} + \delta^0_\mu \delta^0_\nu \right) a^2 V'(\phi) \phi^{(1)} \eqend{,}
\end{splitequation}
where
\begin{equation}
R^\text{flat}_{\mu\nu} = \partial_{(\mu} \partial^\beta h_{\nu)\beta} - \frac{1}{2} \partial^2 h_{\mu\nu} - \frac{1}{2} \partial_\mu \partial_\nu h
\end{equation}
and
\begin{equation}
R^\text{flat} = \eta^{\mu\nu} R^\text{flat}_{\mu\nu} = \partial^\alpha \partial^\beta h_{\alpha\beta} - \partial^2 h
\end{equation}
are the linearised flat-space Ricci tensor and scalar. We did not use the linearised field equation $F^{(1)}$, but we list it for completeness:
\begin{splitequation}
\frac{H a^3}{\phi'} F^{(1)} &= \partial^2 \left( \frac{H a}{\phi'} \phi^{(1)} \right) + H a (n-2\epsilon) \partial_\eta \left( \frac{H a}{\phi'} \phi^{(1)} \right) + \frac{a^2}{\phi'} V'(\phi) \partial_\eta \left( \frac{H a}{\phi'} \phi^{(1)} \right) \\
&\quad+ \frac{H a}{n-2} \left[ H a V'(\phi) + \phi' (n-1-\epsilon) \right] \phi^{(1)} + H a \left( \partial^\rho h_{0\rho} - \frac{1}{2} h' \right) + \frac{1}{2} \frac{H a^3}{\phi'} V'(\phi) h_{00} \eqend{.}
\end{splitequation}
We then find after a long but straightforward calculation
\begin{equation}
C_{\mu\nu} + t_\mu t_\nu g^{\rho\sigma} C_{\rho\sigma} = C_{\mu\nu} + \delta_\mu^0 \delta_\nu^0 \eta^{\rho\sigma} C_{\rho\sigma} = U_{\mu\nu}^{(1)} \eqend{,}
\end{equation}
where $U_{\mu\nu}^{(1)}$ is the first-order part of
\begin{splitequation}
\tilde{U}_{\mu\nu} &\equiv \frac{1}{n-3} \left( - 2 \tilde{u}^\alpha \tilde{u}^\beta \tilde{C}_{\alpha\mu\beta\nu} + 2 \tilde{\mathcal{R}}_{\mu\nu} - \frac{1}{n-2} \tilde{g}_{\mu\nu} \tilde{\mathcal{R}} \right) + \frac{4}{n-1} \tilde{u}_{(\mu} \tilde{a}_{\nu)} \tilde{\nabla}_\alpha \tilde{u}^\alpha \\
&\quad- \frac{2}{n-2} \tilde{u}_{(\mu} \tilde{E}_{\nu)\beta} \tilde{u}^\beta - \frac{4}{n-2} \tilde{u}_\mu \tilde{u}_\nu \tilde{u}^\alpha \tilde{u}^\beta \tilde{E}_{\alpha\beta} + \frac{n-4}{(n-2) (n-3)} \tilde{u}_\mu \tilde{u}_\nu \tilde{\mathcal{R}} \eqend{.}
\end{splitequation}
(We note that the background value of $\tilde{U}_{\mu\nu}$ vanishes.) Thus, we have succeeded in expressing the traceless parts of the tensor $C_{\mu\nu}$ using (perturbed) local geometric tensors in the FLRW background. However, we have been unable to do so for the trace, or, equivalently, for the $00$ components.

In general, the gauge transformation of a first-order perturbation is given by~\eqref{gauge_trafo_lie}
\begin{equation}
\delta T_{\alpha\beta\cdots\gamma}^{(1)} = \mathscr{L}_\xi T_{\alpha\beta\cdots\gamma} \eqend{,}
\end{equation}
which vanishes for arbitrary $\xi$ if and only if the background value vanishes $T_{\alpha\beta\cdots\gamma} = 0$, or is proportional to a combination of $\delta$ tensors, in which case we can redefine $T_{\alpha\beta\cdots\gamma}$ by subtracting these terms without changing the perturbation $T_{\alpha\beta\cdots\gamma}^{(1)}$. Therefore, we can obtain all gauge-invariant linearised observables expressable using (perturbed) local geometric tensors by linearising all geometric tensors which vanish for the FLRW background. A complete set of such tensors has been found recently~\cite{canepa_dappiaggi_khavkine_2017}, but so far no suitable combination of their linearisations has been found.

\bibliography{equivalenceweyl}

\end{document}